\documentclass[12pt,a4paper]{article}

\usepackage{fullpage}
\usepackage[utf8]{inputenc}
\usepackage{amssymb}
\usepackage{amsmath}
\usepackage{amsthm}
\usepackage[round]{natbib}
\usepackage[colorlinks=true,linkcolor=blue,citecolor=blue,pdfborder={0 0 0}]{hyperref}
\usepackage{graphicx}
\usepackage{multirow}
\usepackage{setspace}
\usepackage{enumerate}
\usepackage{color}
\usepackage{booktabs,rotating}
\usepackage{bm}
\usepackage{xcolor,colortbl}

\definecolor{Gray}{gray}{0.9}
\newcolumntype{a}{>{\columncolor{Gray}}c}
\newcolumntype{b}{>{\columncolor{white}}c}

\newcommand{\N}{\mathbb{N}}
\newcommand{\R}{\mathbb{R}}
\newcommand{\dd}{\mathrm{d}}
\newcommand{\Cb}{\mathbb{C}}
\newcommand{\CC}{\mathcal{C}}
\newcommand{\QQ}{\mathcal{Q}}
\newcommand{\RR}{\mathcal{R}}
\renewcommand{\SS}{\mathcal{S}}
\newcommand{\TT}{\mathcal{T}}
\newcommand{\1}{\mathbf{1}}
\newcommand{\pobs}[1]{\hat{\bm #1}}
\newcommand{\ip}[1]{\lfloor #1 \rfloor}

\theoremstyle{definition}
\newtheorem{proc}{Procedure}[section]

\numberwithin{equation}{section}

\title{Some copula inference procedures adapted to the presence of ties}

\author{
  Ivan Kojadinovic\,\footnote{Universit\'e de Pau et des Pays de l'Adour,
Laboratoire de math\'ematiques et de leurs applications,
UMR CNRS 5142, B.P. 1155, 64013 Pau Cedex, France.
{E-mail:} \texttt{ivan.kojadinovic@univ-pau.fr}}
}

\begin{document}
\maketitle

\begin{abstract}
When modeling the distribution of a multivariate continuous random vector using the so-called \emph{copula approach}, it is not uncommon to have ties in the coordinate samples of the available data because of rounding or lack of measurement precision. Yet, the vast majority of existing inference procedures on the underlying copula were both theoretically derived and practically implemented under the assumption of no ties. Applying them nonetheless can lead to strongly biased results. Some of the existing statistical tests can however be adapted to provide meaningful results in the presence of ties. It is the case of some tests of exchangeability, radial symmetry, extreme-value dependence and goodness of fit. Detailed algorithms for computing approximate p-values for the modified tests are provided and their finite-sample behaviors are empirically investigated through extensive Monte Carlo experiments. An illustration on a real-world insurance data set concludes the work.

\medskip

\noindent {\it Keywords:} bootstrap; exchangeability;  extreme-value dependence; goodness of fit; parametric bootstrap; radial symmetry; statistical tests; ties.


\end{abstract}


\section{Introduction}

The copula approach to the modeling of multivariate continuous distributions is increasingly applied in numerous fields such as environmental modeling \citep{SalDeMKotRos07}, quantitative risk management \citep{McNFreEmb15} or econometric modeling \citep{Pat12}, to name a few.

Let $\bm X_1,\dots,\bm X_n$ be independent and identically distributed (i.i.d.) copies of a random vector $\bm X$ with $d$-dimensional cumulative distribution function (c.d.f.) $F$. The use of copulas to model $F$ from $\bm X_1,\dots,\bm X_n$ becomes particularly meaningful when the $d$ univariate marginal c.d.f.s (\emph{margins} for short) $F_1,\dots,F_d$ associated with $F$ are assumed continuous. Indeed, in that case, the copula $C$ (merely a multivariate c.d.f.\ with standard uniform univariate margins) associated with $F$ through the well-known representation
\begin{equation}
\label{eq:F}
F(\bm x) = C\{F_1(x_1),\dots,F_d(x_d) \}, \qquad \bm x \in \R,
\end{equation}
is unique and is given by
\begin{equation}
\label{eq:C}
C(\bm u) = F \{F_1^{-1} (u_1),\dots,F_d^{-1}(u_d) \}, \qquad \bm u \in [0,1]^d,
\end{equation}
in terms of the quantile functions (generalized inverses) $F_1^{-1},\dots,F_d^{-1}$ associated with $F_1,\dots,F_d$, respectively; see \citet{Skl59} and, for instance, \cite{Rus09}. To obtain a parametric estimate of $F$ with~\eqref{eq:F} in mind, a practitioner needs to model $F_1, \dots, F_d$ by appropriate univariate parametric families, and $C$ by an adequate parametric copula family. This work is concerned with the latter step only.

There exists a large number of parametric copula families that belong to broader classes such as \emph{extreme-value} copulas \citep[see, e.g,][]{GudSeg10}, \emph{Archimedean} copulas \citep[see, e.g,][]{Nel06} or \emph{elliptical} copulas \citep[see, e.g.,][Chap.~7]{McNFreEmb15}. To help guide the choice of the most appropriate parametric copula family for the data at hand, many inference procedures, mostly taking the form of statistical tests, were proposed in the literature. 
For instance, tests of \emph{extreme-value dependence} were proposed by \cite{GhoKhoRiv98}, \cite{BenGenNes09}, \cite{BucDetVol11}, \cite{KojSegYan11} and \cite{CorGenNes14}, among others, bivariate tests of \emph{exchangeability} were proposed by \cite{GenNesQue12} and \cite{KojYan12b}, bivariate tests of \emph{radial symmetry} were investigated in \cite{GenNes14} while goodness-of-fit tests were studied for instance in \cite{GenRemBea09}, \cite{Ber09} and \cite{KojYan11} \citep[see, e.g.,][for a recent review]{Fer13}.

The vast majority of existing tests on the unknown copula $C$ were theoretically investigated and practically implemented under the assumption of no ties in the coordinate samples of the available i.i.d.\ data $\bm X_1,\dots,\bm X_n$. Indeed, if the $d$ univariate margins of $F$ are continuous, ties cannot occur in the component series. Yet, because of rounding or measurement precision issues, it is not uncommon to have ties in real-world data sets, even if the underlying random phenomenon to be modeled is truly continuous.

Several practical studies such as those in \cite{KojYan10}, \cite{GenNesRup11} or \cite{PapDurSal16} highlight the fact that the presence of ties in the coordinate samples can strongly bias the results of the aforementioned inference procedures on $C$. Additional empirical evidence will be provided in the forthcoming sections.

The aim of this work is to propose versions of some of the aforementioned tests that are adapted to the presence of ties in the component series of $\bm X_1,\dots,\bm X_n$. In particular, modifications of the bivariate tests of exchangeability of \cite{GenNesQue12} and \cite{KojYan12b}, of the test of radial symmetry of \cite{GenNes14}, of the bivariate tests of extreme-value dependence of \cite{GhoKhoRiv98} and \cite{BenGenNes09}, and of the parametric bootstrap-based goodness-of-fit tests of \cite{GenRemBea09} are suggested and empirically investigated in a large number of Monte Carlo experiments. Unfortunately, given the difficulty of the underlying theoretical setting \citep[see, e.g.,][]{GenNes07,GenNesRem14}, no theoretical results on the asymptotic validity of the modified tests are provided.

The second section details the data generating mechanism used in the simulations. The tie-adapted tests are described and empirically studied in Sections~\ref{sec:exch}--\ref{sec:gof}. An illustration on a real-world insurance data set is provided in Section~\ref{sec:illus} while the last section concludes. Notice that all the tests studied in this work are implemented in the package \texttt{copula} \citep{copula} for the \textsf{R} statistical environment \citep{Rsystem}, making all the numerical experiments presented in the paper fully reproducible.

\section{Data generating mechanism}

To empirically investigate the finite-sample behaviors of the tie-adapted tests to be presented in the forthcoming sections, we carried out numerous Monte Carlo experiments. Given a $d$-dimensional copula $C$, a sample size~$n$ and two discretization parameters $k \in \N$, $k \geq 1$, and $t \in \{0.5,1,2\}$, samples were generated using the following procedure:

\begin{proc}[Data generating mechanism]
\label{proc:dgm}
$\strut$
\begin{enumerate}[(i)]
\item Generate a random sample $\bm U_1,\dots,\bm U_n$ from $C$.
\item Form the $k$ bins $(a_i,a_{i+1}]$, $i \in \{0,\dots,k-1\}$, where $a_i = i^t / k^t$, $i \in \{0,\dots,k\}$.
\item For all $j \in \{1,\dots,d\}$, replace each observation in the coordinate sample $U_{1j},\dots,U_{nj}$ by the center of the bin to which it belongs.
\end{enumerate}
\end{proc}
In the sequel, we adopt the convention that the setting $k = \infty$ corresponds to no discretization. Notice that taking $t \in \{0.5,2\}$ results in bins of different lengths.

\section{Tests of exchangeability}
\label{sec:exch}

A $d$-dimensional copula $C$ is said to be \emph{exchangeable} if, for any $\bm u \in [0,1]^d$ and any permutation $\pi$ on $\{1,\dots,d\}$, $C(\bm u) = C(u_{\pi(1)},\dots,u_{\pi(d)})$. Many of the parametric copula families used in practice are exchangeable. Various mechanisms can however be used to construct non-exchangeable families from exchangeable ones \citep[see, e.g.,][for a review]{GenNes13}. One construction principle of particular interest is known as \emph{Khoudraji's device}. It was initially proposed by \cite{Kho95} and is further discussed in \cite{GenGhoRiv98} and \cite{GenNes13}. It turns out to be a particular case of the more general construction principal proposed by \cite{Lie08}.

Given two $d$-dimensional copulas $C_1$ and $C_2$, and a \emph{shape vector}
$\bm s \in [0,1]^d$, \emph{Khoudraji's device} creates a new copula $D$ as
\begin{equation}
  \label{eq:khoudraji}
D(\bm u) =   C_1(\bm u^{\1-\bm s}) C_2(\bm u^{\bm s}), \qquad \bm u \in [0,1]^d,
\end{equation}
with the convention that $\bm u^{\bm v} = (u_1^{v_1},\dots,u_d^{v_d})$ for all
$\bm u, \bm v \in [0,1]^d$.

To decrease the number of candidate parametric families when carrying out inference on the unknown copula $C$, it is of interest to assess from $\bm X_1,\dots,\bm X_n$ whether $C$ is exchangeable. A formal test of exchangeability was studied in the bivariate case  by \cite{GenNesQue12} as a particular case of the one proposed by \cite{RemSca09}. It is based on the statistic
\begin{align}
\nonumber
\RR_{n,C} &= n \int_{[0,1]^2} \{ C_n(u_1,u_2) -  C_n(u_2,u_1) \}^2 \dd C_n(\bm{u}) \\
\label{eq:RnC}
&= \sum_{i=1}^n \{ C_n(\hat U_{i1}, \hat U_{i2}) -  C_n(\hat U_{i2}, \hat U_{i1}) \}^2,
\end{align}
where $C_n$ is a consistent nonparametric estimator of the unknown copula $C$ based on $\bm X_1,\dots,\bm X_n$ that we shall refer to as the \emph{empirical copula}. For arbitrary $d \geq 2$, it is defined as
\begin{equation}
\label{eq:empcop}
C_n(\bm u) = \frac{1}{n} \sum_{i=1}^n \1 ( \pobs{U}_i \leq \bm u ), \qquad \bm u \in [0,1]^d,
\end{equation}
where inequalities between vectors are to be understood componentwise,
\begin{equation}
\label{eq:pobs}
\pobs{U}_i = \frac{1}{n+1}(R_{i1},\dots,R_{id}), \qquad i \in \{1,\dots,n\},
\end{equation}
and, for any $j \in \{1,\dots,d\}$, $R_{1j},\dots,R_{nj}$ are the ranks computed from $X_{1j},\dots,X_{nj}$.

Before proceeding further and under the assumption of no ties in the components samples of $\bm X_1,\dots,\bm X_n$, let us make a few remarks:
\begin{itemize}
\item Recall that $R_{ij} = n F_{n,j}(X_{ij})$, $i \in \{1,\dots,n\}$, $j \in \{1,\dots,d\}$, where $F_{n,j}$ is the empirical c.d.f.\ computed from $X_{1j},\dots,X_{nj}$.
\item Starting from~\eqref{eq:C} and applying the plug-in principle, a seemingly more natural nonparametric estimator of $C$, considered for instance in \citet{Deh79,Deh81}, is
\begin{equation}
\label{eq:empcop2}
\tilde C_n(\bm u) = F_n\{F_{n,1}^{-1}(u_1), \dots, F_{n,d}^{-1}(u_d) \}, \qquad \bm u \in [0,1]^d,
\end{equation}
where $F_n$ is the multivariate empirical c.d.f.\ computed from $\bm X_1,\dots,\bm X_n$ and, for any $j \in \{1,\dots,d\}$, $F_{n,j}^{-1}$ is the generalized inverse of the univariate empirical c.d.f.~$F_{n,j}$.

\item The difference between $C_n$ and $\tilde C_n$ is of the order $1/n$ but the former is substantially simpler to compute.

\item If the division by $n+1$ in~\eqref{eq:pobs} were replaced by division by $n$, $C_n$ and $\tilde C_n$ would coincide on the set $\{(i_1/n,\dots,i_d/n) : i_1,\dots,i_d \in \{1,\dots,n\}\}$ and thus $C_n$ in the expression of $\RR_{n,C}$ in~\eqref{eq:RnC} could be replaced by $\tilde C_n$.

\item The division by $n+1$ in~\eqref{eq:pobs} is carried out in this work with maximum pseudo-likelihood estimation \citep[see][]{GenGhoRiv95} and nonparametric estimation of the Pickands dependence function in mind \citep[see][]{GenSeg09}; see also later in this section and Section~\ref{sec:gof}.

\end{itemize}

To attempt to extend the test of exchangeability based on $\RR_{n,C}$ to the presence of ties in the component samples of $\bm X_1,\dots,\bm X_n$, we first need to decide how to compute the ranks involved in~\eqref{eq:pobs} in that case. A first sensible approach would consist of considering \emph{maximal} ranks. It is only if this definition is used that one recovers in the presence of ties the well-known fact in the absence of ties that $R_{ij} = n F_{n,j}(X_{ij})$, $i \in \{1,\dots,n\}$, $j \in \{1,\dots,d\}$. An alternative approach, common in nonparametric statistics, would be to use \emph{average} ranks, also called \emph{midranks}; see, for instance, \citet{Agr02, Agr10}, or the \textsf{R} function \texttt{rank} which computes average ranks by default. Roughly speaking, if average (resp.\ maximal) ranks are use, tied observations are assigned the average (resp.\ the maximum) of the ranks they would obtain if there were no ties. In the absence of ties, as desired, all definitions lead to the same result. In the rest of this work, we shall use either maximal ranks or average ranks for reasons that we will always be explicitly stated.

Going back to the test of exchangeability of \cite{GenNesQue12}, our numerical experiments indicate that, in the presence of ties, the type of ranks (maximal or average) used in~\eqref{eq:pobs} when computing $\RR_{n,C}$ in~\eqref{eq:RnC} does not seem to have much influence on the results. We have thus arbitrarily decided to compute $\RR_{n,C}$ from scaled average ranks.

An alternative test of exchangeability was proposed by \cite{KojYan12b}, initially, under the additional assumption that the unknown bivariate copula $C$ is an \emph{extreme-value} copula \citep[see, e.g.,][for an overview of the main characterizations and properties of such copulas]{GudSeg10}. The test statistic is defined as
\begin{equation}
\label{eq:RnA}
\RR_{n,A} = n \int_{[0,1]} \{ A_n(t) -  A_n(1-t) \}^2 \dd t,
\end{equation}
where $A_n$ is the rank-based version of the \emph{Cap\'era\`a--Foug\`eres--Genest estimator} \citep{CapFouGen97} of the \emph{Pickands dependence function} \citep{Pic81} associated with $C$. The estimator $A_n$ is defined in Eq.~(2.3) of \cite{GenSeg09}, where its most important theoretical properties are established. For our purpose, it is sufficient to keep in mind that $A_n$ solely depends on the bivariate scaled ranks $\pobs{U}_1,\dots,\pobs{U}_n$ defined in~\eqref{eq:pobs}, making the statistic $\RR_{n,A}$ rank-based. Our Monte-Carlo experiments suggest that, in the presence of ties, $A_n$ should be computed from scaled average ranks.

The latter test remains actually meaningful when $C$ belongs to the larger class of bivariate copulas that are \emph{left-tail decreasing} (LTD) in both variables \citep[see, e.g.,][Section 5.2.2]{Nel06}. Note that, from \citet[Exercise 5.35]{Nel06}, a bivariate copula $C$ is LTD in both arguments if and only if, for any $0 < u \leq u' \leq 1$ and $0 < v \leq v' \leq 1$,
$$
\frac{C(u,v)}{uv} \geq \frac{C(u',v')}{u'v'}.
$$
As shown by \cite{Gar00}, bivariate extreme-value copulas are LTD in both variables but so are the most popular bivariate copulas with positive dependence such as the Clayton, Frank, normal and Plackett.

For the computation of approximate p-values, both the test based on $\RR_{n,C}$ and the one based on $\RR_{n,A}$ rely on (an adaptation of) the \emph{multiplier} resampling scheme proposed by \cite{RemSca09}. While, in the absence of ties, the two tests do not appear to be too liberal (that is, they do not seem to reject the null hypothesis too often when it is true), it is not the case anymore when the amount of ties in the coordinate samples of $\bm X_1,\dots,\bm X_n$ becomes non negligible. This will be illustrated later in this section when reporting the results of certain Monte Carlo experiments.

Let $\RR_n$ stand for $\RR_{n,C}$ in~\eqref{eq:RnC} or $\RR_{n,A}$ in~\eqref{eq:RnA}. In order to obtain a version of the test based on $\RR_n$ adapted to ties, we propose the following bootstrap procedure:

\begin{proc}[Approximate p-value for $\RR_n$ in the presence of ties]
\label{proc:exch}
$\strut$
\begin{enumerate}
\item For $j\in\{1,2\}$, compute the average ranks $R_{1j},\dots,R_{nj}$ of $X_{1j},\dots,X_{nj}$ and set $S_{ij} = R_{\sigma_j(i), j}$, $i \in \{1,\dots,n\}$, where $\sigma_j$ is a permutation on $\{1,\dots,n\}$ such that $S_{1j} = R_{\sigma_j(1), j} \leq \dots \leq S_{nj} = R_{\sigma_j(n), j}$.
\item Compute $\pobs{U}_1,\dots,\pobs{U}_n$ from the average ranks computed in Step~1 using~\eqref{eq:pobs} and, then, compute  $\RR_n$ from $\pobs{U}_1,\dots,\pobs{U}_n$.
\item For some large integer $N$, repeat the following steps for every $k \in \{1,\dots,N\}$:
  \begin{enumerate}
  \item Generate $n$ random permutations $\pi_1^{(k)},\dots,\pi_n^{(k)}$ on $\{1,2\}$ and form the sample
    $$
    \pobs{V}_i^{(k)} = (\hat U_{i,\pi_i^{(k)}(1)},\hat U_{i,\pi_i^{(k)}(2)}), \qquad i \in \{1,\dots,n\}.
    $$
  \item Set $\pobs{W}_i^{(k)} = \pobs{V}_i^{(k)}$, $i \in \{1,\dots,n\}$, and then, for $j \in \{1,2\}$:
    \begin{itemize}
    \item[-] find a permutation $\rho^{(k)}_j$ on $\{1,\dots,n\}$ such that $\hat W_{\rho^{(k)}_j(1),j}^{(k)} \leq \dots \leq \hat W_{\rho^{(k)}_j(n),j}^{(k)}$,
    \item[-] set $\hat W_{\rho^{(k)}_j(i),j}^{(k)} = \hat V_{\ip{S_{ij}},j}^{(k)}$, $i \in \{1,\dots,n\}$,
    \end{itemize}
    where $\ip{\cdot}$ is the floor function.
\item Form the $k$th bootstrap sample as
$$
\pobs{U}_i^{(k)} = \frac{1}{n+1} (R_{i1}^{(k)},R_{i2}^{(k)}), \qquad i \in \{1,\dots,n\},
$$
where, for any $j \in \{1,2\}$, $R_{1j}^{(k)},\dots,R_{nj}^{(k)}$ are the average ranks computed from $\hat W_{1j}^{(k)},\dots,\hat W_{nj}^{(k)}$.

\item Let $\RR_n^{(k)}$ stand for the version of $\RR_n$ computed from $\pobs{U}_1^{(k)},\dots,\pobs{U}_n^{(k)}$.
  \end{enumerate}
\item An approximate p-value for the test is given by
\begin{equation}
\label{eq:pval}
\frac{1}{N+1} \sum_{k=1}^N \{\1(\RR_{n}^{(k)} \geq \RR_n) + 0.5\}.
\end{equation}
\end{enumerate}
\end{proc}

Let us comment on the above procedure:
\begin{itemize}
\item The aim of Step~3~(a) is to ``break the non-echangeability'', if any, in $\pobs{U}_1,\dots,\pobs{U}_n$.
\item The aim of Steps 3~(b) and 3~(c) is that each bootstrap sample $\pobs{U}_1^{(k)},\dots,\pobs{U}_n^{(k)}$ has similar marginal empirical c.d.f.s as $\pobs{U}_1,\dots,\pobs{U}_n$ in~\eqref{eq:pobs}. The key ingredient is Step~3~(b) which exploits an idea from \citet[Section 5.2]{BucKoj15} that should be credited to the first author of the latter reference. Roughly speaking, Step~3~(b) imposes on the sample $\pobs{V}_1^{(k)},\dots,\pobs{V}_n^{(k)}$ resulting from Step~3~(a) the ``tie structure'' found in the coordinate samples of $\bm X_1,\dots,\bm X_n$. Although this is not true in general, it helps to think of the latter as consisting of applying on the coordinate samples of $\pobs{V}_1^{(k)},\dots,\pobs{V}_n^{(k)}$ a certain empirical version of the quantile transformation. As a consequence of Steps 3~(b) and 3~(c), $\RR_n, \RR_n^{(1)}, \dots, \RR_n^{(N)}$ are computed from samples with similar marginal empirical c.d.f.s.
\item The slight modification in~\eqref{eq:pval} of the classical formula $N^{-1} \sum_{k=1}^N \1(\RR_n^{(k)} \geq \RR_n)$ is used to ensure that the p-value is in the open interval $(0,1)$ so that transformations by quantile functions of continuous distributions are always well-defined.
\item Some thought reveals that the procedure remains meaningful even if there are no ties in the coordinate samples of $\bm X_1,\dots,\bm X_n$.
\end{itemize}

As we continue, to refer to the original tests based on~\eqref{eq:RnC} and~\eqref{eq:RnA}, we shall use the expressions \emph{the test based on $\RR_{n,C}$} and \emph{the test based on $\RR_{n,A}$}, respectively, while to refer to the tests adapted to ties, we shall write \emph{the test based on $\RR_{n,C}'$} and \emph{the test based on $\RR_{n,A}'$}, respectively.

To investigate the levels of the tests based on $\RR_{n,C}$, $\RR_{n,C}'$, $\RR_{n,A}$ and $\RR_{n,A}'$ empirically, we generated 1000 samples using Procedure~\ref{proc:dgm} for $n \in \{50,100,200\}$, $k \in \{10,20,50,\infty\}$, $t \in \{0.5,1,2\}$ and $C$ either the bivariate Clayton, Gumbel--Hougaard, Frank, normal or Plackett copula with a Kendall's tau of $\tau \in \{0,0.25,0.5,0.75\}$. For each combination of $C$, $\tau$, $n$, $k$ and $t$, the tests were then carried out at the 5\% significance level and approximate p-values were computed from $N=1000$ multiplier or bootstrap replicates. A subset of the obtained rejection percentages for $k \in \{10,\infty\}$ and $t = 1$ is reported in Table~\ref{exchH0}.

\begin{table}[t!]
\centering
\caption{Percentages of rejection of the null hypothesis of exchangeability computed from 1000 samples of size $n \in \{50, 100, 200\}$ generated using Procedure~\ref{proc:dgm} with $C$ the Clayton or Gumbel--Hougaard copula with a Kendall's tau of $\tau \in \{0,0.25,0.5,0.75\}$, $k \in \{10,\infty\}$ and $t=1$.} 
\label{exchH0}
\begin{tabular}{rrlrrrrrrrr}
  \hline
  \multicolumn{3}{c}{} & \multicolumn{4}{c}{Cl} & \multicolumn{4}{c}{GH} \\ \cmidrule(lr){4-7} \cmidrule(lr){8-11} $\tau$ & $n$ & $k$ & $\RR_{n,C}$ & $\RR_{n,C}'$ & $\RR_{n,A}$ & $\RR_{n,A}'$& $\RR_{n,C}$ & $\RR_{n,C}'$ & $\RR_{n,A}$ & $\RR_{n,A}'$ \\ \hline
0.00 & 50 & $\infty$ & 4.9 & 3.7 & 6.5 & 6.2 & 5.2 & 3.4 & 7.0 & 6.3 \\ 
   &  & 10 & 100.0 & 1.1 & 6.8 & 5.9 & 100.0 & 0.7 & 6.1 & 5.5 \\ 
   & 100 & $\infty$ & 3.5 & 3.2 & 5.3 & 5.5 & 3.1 & 2.7 & 4.3 & 4.1 \\ 
   &  & 10 & 100.0 & 2.6 & 5.5 & 4.7 & 100.0 & 2.2 & 6.3 & 5.2 \\ 
   & 200 & $\infty$ & 3.5 & 3.7 & 5.0 & 5.6 & 4.4 & 4.2 & 5.5 & 5.7 \\ 
   &  & 10 & 100.0 & 3.9 & 5.1 & 4.4 & 100.0 & 3.7 & 4.3 & 3.9 \\ 
  0.25 & 50 & $\infty$ & 3.0 & 0.8 & 5.3 & 5.5 & 4.0 & 1.8 & 4.8 & 6.0 \\ 
   &  & 10 & 100.0 & 0.4 & 6.0 & 5.6 & 100.0 & 0.6 & 4.9 & 5.7 \\ 
   & 100 & $\infty$ & 2.2 & 2.1 & 5.4 & 6.1 & 3.1 & 2.1 & 3.9 & 4.1 \\ 
   &  & 10 & 100.0 & 1.4 & 6.3 & 6.1 & 100.0 & 0.9 & 5.4 & 5.3 \\ 
   & 200 & $\infty$ & 3.8 & 3.9 & 4.7 & 5.0 & 2.3 & 2.5 & 5.2 & 5.6 \\ 
   &  & 10 & 100.0 & 3.4 & 5.6 & 4.9 & 100.0 & 2.2 & 4.5 & 3.7 \\ 
  0.50 & 50 & $\infty$ & 2.0 & 0.8 & 3.3 & 5.6 & 3.2 & 0.9 & 3.0 & 6.3 \\ 
   &  & 10 & 100.0 & 0.4 & 4.9 & 5.7 & 100.0 & 0.4 & 5.5 & 5.4 \\ 
   & 100 & $\infty$ & 1.7 & 1.5 & 4.3 & 5.7 & 2.2 & 1.7 & 2.4 & 5.4 \\ 
   &  & 10 & 100.0 & 2.1 & 4.3 & 4.3 & 100.0 & 1.2 & 6.7 & 5.4 \\ 
   & 200 & $\infty$ & 2.5 & 2.6 & 4.7 & 5.3 & 2.2 & 2.5 & 3.6 & 4.9 \\ 
   &  & 10 & 100.0 & 2.7 & 4.3 & 4.0 & 100.0 & 2.4 & 6.6 & 5.2 \\ 
  0.75 & 50 & $\infty$ & 2.8 & 0.1 & 1.9 & 6.6 & 2.4 & 0.3 & 0.3 & 7.7 \\ 
   &  & 10 & 100.0 & 0.3 & 3.0 & 4.2 & 100.0 & 0.3 & 5.8 & 3.6 \\ 
   & 100 & $\infty$ & 2.3 & 0.7 & 2.1 & 6.6 & 1.7 & 0.2 & 1.0 & 6.5 \\ 
   &  & 10 & 100.0 & 0.8 & 4.5 & 3.2 & 100.0 & 1.1 & 14.2 & 5.4 \\ 
   & 200 & $\infty$ & 2.2 & 0.7 & 3.5 & 5.7 & 1.5 & 1.0 & 1.2 & 4.6 \\ 
   &  & 10 & 100.0 & 2.3 & 6.2 & 3.8 & 100.0 & 1.9 & 18.3 & 3.2 \\ 
   \hline
\end{tabular}
\end{table}

As one can see, the tests do not seem too liberal when there is no discretization ($k=\infty$). For $k=10$, however, the test based on $\RR_{n,C}$ appears to reject the null hypothesis of exchangeability almost always. The test based on $\RR_{n,A}$ seems much more robust against ties and it is only when the dependence is strong ($\tau = 0.75$) that its levels sometimes appear to be substantially larger than the 5\% nominal level. The results for the other copula families (not reported), and the other values of $k$ and $t$, are not qualitatively different. Unlike the tests based on $\RR_{n,C}$ and $\RR_{n,A}$, the tests adapted to ties were never observed to be too liberal. The test based on $\RR_{n,C}'$ appears however too conservative, in particular when the dependence is strong ($\tau = 0.75$), although the agreement with the 5\% nominal level seems to improve as $n$ increases.

\begin{table}[t!]
\centering
\caption{Percentages of rejection of the null hypothesis of exchangeability computed from 1000 samples of size $n \in \{100, 200\}$ generated from the copula in~\eqref{eq:khoudraji} with $C_1$ the independence copula, $C_2$ the normal copula with a Kendall's tau of 0.75, $s_1 \in \{0.2,0.4,0.6,0.8\}$, $s_2 = 0.95$, and without additional discretization.} 
\label{exchH1kinf}
\begin{tabular}{rrrrrr}
  \hline
  $s_1$ & $n$ & $\RR_{n,C}$ & $\RR_{n,C}'$ & $\RR_{n,A}$ & $\RR_{n,A}'$ \\ \hline
0.2 & 100 & 43.6 & 40.7 & 73.1 & 73.9 \\ 
   & 200 & 80.9 & 80.8 & 97.9 & 98.1 \\ 
  0.4 & 100 & 77.1 & 72.6 & 92.1 & 94.5 \\ 
   & 200 & 99.1 & 99.2 & 100.0 & 100.0 \\ 
  0.6 & 100 & 71.6 & 62.4 & 80.4 & 81.5 \\ 
   & 200 & 97.6 & 96.7 & 98.0 & 97.9 \\ 
  0.8 & 100 & 21.4 & 14.3 & 33.4 & 39.3 \\ 
   & 200 & 51.1 & 44.9 & 57.6 & 58.3 \\ 
   \hline
\end{tabular}
\end{table}
\begin{table}[t!]
\centering
\caption{Percentages of rejection of the null hypothesis of exchangeability computed from 1000 samples of size $n \in \{100, 200\}$ generated from the copula in~\eqref{eq:khoudraji} with $C_1$ the independence copula, $C_2$ the normal copula with a Kendall's tau of 0.75, $\bm s = (0.2,0.95)$, and further discretized as explained in Procedure~\ref{proc:dgm} with $k \in \{10,20,50\}$ and $t \in \{0.5,1,2\}$.} 
\label{exchH1allk}
\begin{tabular}{rrrrrr}
  \hline
  \multicolumn{2}{c}{} & \multicolumn{2}{c}{$n=100$} & \multicolumn{2}{c}{$n=200$} \\ \cmidrule(lr){3-4} \cmidrule(lr){5-6} $k$ & $t$ & $\RR_{n,C}'$ & $\RR_{n,A}'$ & $\RR_{n,C}'$ & $\RR_{n,A}'$ \\ \hline
10 & 1.0 & 19.8 & 60.6 & 46.5 & 89.9 \\ 
   & 2.0 & 15.6 & 45.5 & 33.0 & 76.2 \\ 
   & 0.5 & 7.5 & 62.1 & 25.6 & 91.9 \\ 
  20 & 1.0 & 28.4 & 71.7 & 65.1 & 95.4 \\ 
   & 2.0 & 21.6 & 61.6 & 50.0 & 89.8 \\ 
   & 0.5 & 11.1 & 67.5 & 39.8 & 94.9 \\ 
  50 & 1.0 & 37.7 & 75.5 & 76.0 & 96.4 \\ 
   & 2.0 & 34.1 & 71.9 & 75.8 & 96.8 \\ 
   & 0.5 & 22.7 & 74.2 & 48.8 & 97.4 \\ 
   \hline
\end{tabular}
\end{table}

An additional issue of practical interest is related to the observation that the adapted bootstrap given in Procedure~\ref{proc:exch} remains meaningful even if there are no ties in the coordinate samples of $\bm X_1,\dots,\bm X_n$. To investigate if the use of the bootstrap procedure leads to a loss of power when there is no discretization, we carried out the tests on samples generated from non-exchangeable copulas constructed using Khoudraji's device. The copula $C_1$ in~\eqref{eq:khoudraji} was taken to be the independence copula, the copula $C_2$ was taken to be either the Clayton, Gumbel--Hougaard, Franck, normal or Plackett copula with a Kendall's tau of 0.75, the second shape parameter $s_2$ was set to 0.95, while the first shape parameter $s_1$ was taken in $\{0.2,0.4,0.6,0.8\}$ so that the resulting copula $D$ displays various degrees of asymmetry with respect to the first diagonal. The rejection percentages of the null hypothesis of exchangeability when $C_2$ is the normal copula are reported in Table~\ref{exchH1kinf}. As one can notice, the test based on $\RR_{n,C}$ (resp. $\RR_{n,A}'$) might be slightly more powerful than the one based on $\RR_{n,C}'$ (resp. $\RR_{n,A}$).

In a final experiment, we investigated the influence of $k$ and $t$ on the power of the tie-adapted tests. To do so, samples of size $n \in \{100, 200\}$ were generated from the copula in~\eqref{eq:khoudraji} with $C_1$ the independence copula, $C_2$ either the Clayton, Gumbel--Hougaard, Franck, normal or Plackett copula with a Kendall's tau of 0.75, $s_2 = 0.95$ and $s_1 \in \{0.2,0.4,0.6,0.8\}$, and were further discretized as explained in Procedure~\ref{proc:dgm} with $k \in \{10,20,50\}$ and $t \in \{0.5,1,2\}$. Table~\ref{exchH1allk}  reports the rejection percentages when $C_2$ is the normal copula and $s_1=0.2$. Overall, for fixed $n$ and $t$, the power increases with $k$. In other words, as could have been expected, the smaller the amount of ties, the more powerful the adapted tests.

\section{Tests of radial symmetry}
\label{sec:radSym}

Let $C$ be a $d$-dimensional copula and let $\bm U$ be a random vector with c.d.f.\ $C$. The c.d.f.\ of the random vector $\bm 1 - \bm U$ is called the \emph{survival copula} of $C$ and is classically denoted by $\bar C$. A copula $C$ is said to be \emph{radially symmetric} if it is equal to its survival copula, that is, if $C = \bar C$.

Some copula families, such as the normal or the Frank, are radially symmetric, some others, such as the Clayton or the Gumbel--Hougaard, are not. As for exchangeability, it is thus of strong practical interest to test for radial symmetry when carrying out copula inference. Starting from the above definition of radial symmetry, a natural test statistic is
\begin{align}
  \label{eq:Qn}
  \QQ_n = n \int_{[0,1]^d} \{ C_n(\bm u) -  \bar C_n(\bm u) \}^2 \dd C_n(\bm{u}) = \sum_{i=1}^n \{ C_n(\pobs{U}_i) -  \bar C_n(\pobs{U}_i) \}^2,
\end{align}
where $C_n$ is the empirical copula of $\bm X_1,\dots,\bm X_n$ defined in~\eqref{eq:empcop} and $\bar C_n$ is the empirical copula of $-\bm X_1,\dots,-\bm X_n$. The test based on $\QQ_n$ was studied in the bivariate case by \cite{GenNes14} and relies on a multiplier resampling scheme à la \cite{RemSca09}.

In the presence of ties, it seems sensible to compute $\QQ_n$ from~\eqref{eq:pobs} based on average ranks, in which case, as in the absence of ties, $\bar C_n$ is simply the empirical c.d.f.\ of the sample $\bm{1} - \pobs{U}_1,\dots, \bm{1} - \pobs{U}_n$. By analogy with Procedure~\ref{proc:exch}, a tie-adapted multivariate version of the test can then be carried out as follows:

\begin{proc}[Approximate p-value for $\QQ_n$ in the presence of ties]
\label{proc:radSym}
$\strut$
\begin{enumerate}
\item For $j\in\{1,\dots,d\}$, compute the average ranks $R_{1j},\dots,R_{nj}$ of $X_{1j},\dots,X_{nj}$ and set $S_{ij} = R_{\sigma_j(i), j}$, $i \in \{1,\dots,n\}$, where $\sigma_j$ is a permutation on $\{1,\dots,n\}$ such that $S_{1j} = R_{\sigma_j(1), j} \leq \dots \leq S_{nj} = R_{\sigma_j(n), j}$.
\item Compute $\pobs{U}_1,\dots,\pobs{U}_n$ from the average ranks computed in Step~1 using~\eqref{eq:pobs} and, then, compute  $\QQ_n$ from $\pobs{U}_1,\dots,\pobs{U}_n$.
\item For some large integer $N$, repeat the following steps for every $k \in \{1,\dots,N\}$:
  \begin{enumerate}
  \item Generate $n$ independent copies $Z_1^{(k)},\dots,Z_n^{(k)}$ of a Bernoulli random variable with parameter 0.5 and form the sample
    $$
    \pobs{V}_i^{(k)} = (1 - Z_i^{(k)}) \pobs{U}_i + Z_i^{(k)} ( \bm 1 - \pobs{U}_i), \qquad i \in \{1,\dots,n\}.
    $$

  \item Set $\pobs{W}_i^{(k)} = \pobs{V}_i^{(k)}$, $i \in \{1,\dots,n\}$, and then, for $j \in \{1,\dots,d\}$:
    \begin{itemize}
    \item[-] find a permutation $\rho^{(k)}_j$ on $\{1,\dots,n\}$ such that $\hat W_{\rho^{(k)}_j(1),j}^{(k)} \leq \dots \leq \hat W_{\rho^{(k)}_j(n),j}^{(k)}$,
    \item[-] set $\hat W_{\rho^{(k)}_j(i),j}^{(k)} = \hat V_{\ip{S_{ij}},j}^{(k)}$, $i \in \{1,\dots,n\}$.
    \end{itemize}

\item Form the $k$th bootstrap sample as
$$
\pobs{U}_i^{(k)} = \frac{1}{n+1} (R_{i1}^{(k)},\dots,R_{id}^{(k)}), \qquad i \in \{1,\dots,n\},
$$
where, for any $j \in \{1,\dots,d\}$, $R_{1j}^{(k)},\dots,R_{nj}^{(k)}$ are the average ranks computed from $\hat W_{1j}^{(k)},\dots,\hat W_{nj}^{(k)}$.

\item Let $\QQ_n^{(k)}$ stand for the version of $\QQ_n$ computed from $\pobs{U}_1^{(k)},\dots,\pobs{U}_n^{(k)}$.
  \end{enumerate}
\item An approximate p-value for the test is given by $(N+1)^{-1} \sum_{k=1}^N \{\1(\QQ_{n}^{(k)} \geq \QQ_n) + 0.5\}$.
\end{enumerate}
\end{proc}

Unsurprisingly, the above procedure is very similar to Procedure~\ref{proc:exch} providing tie-adapted versions of the tests of exchangeability considered in Section~\ref{sec:exch}. The main difference lies in Step~3~(a), which, instead of ``breaking the non-echangeability'', ``breaks the radial asymmetry'' in $\pobs{U}_1,\dots,\pobs{U}_n$, if any. Notice that if there are no ties in the coordinate samples of $\bm X_1,\dots,\bm X_n$, the aforementioned procedure can be simplified by removing Steps~1 and~3~(b) provided the average ranks in Step~3~(c) are computed from the coordinate samples of $\pobs{V}_1^{(k)},\dots,\pobs{V}_n^{(k)}$ obtained in Step~3~(a). In the sequel, we shall use the expression \emph{the test based on $\QQ_n$} to designate the test based on the latter simplification, while we shall talk about \emph{the test based on $\QQ_n'$} to refer to the tie-adapted test based on the full procedure as given above.

To empirically assess whether the tests based on $\QQ_n$ and $\QQ_n'$ hold their levels, we generated 1000 samples using Procedure~\ref{proc:dgm} for $n \in \{50,100,200\}$, $k \in \{10,50,\infty\}$, $t=1$ and $C$ either the 2-, 3- or 4-dimensional exchangeable normal or Student $t_4$ copula such that its bivariate margins have a Kendall's tau of $\tau \in \{0,0.25,0.5,0.75\}$. The rejection percentages are given in Table~\ref{radSymH0}. By considering the lines corresponding to no discretization ($k=\infty$), one can see that the empirical levels of the two tests are reasonably close to the 5\% nominal level, except maybe in the case of strong dependence ($\tau=0.75$), the agreement however improving as $n$ increases. For $k \in \{10, 50\}$, the test based on $\QQ_n$ becomes way too liberal, thereby empirically confirming that Steps~1 and~3~(b) in Procedure~\ref{proc:radSym} are crucial. On the contrary, the test based on $\QQ_n'$ becomes too conservative in the presence of ties, although the situation seems overall to improve as $n$ increases. A similar experiment was carried out by considering bivariate Frank and Plackett copulas instead and confirmed the appropriate behavior of the test based on $\QQ_n'$ in that case. The only setting in which inflated empirical levels were observed for the test based on $\QQ_n'$ (as well as for the test based on $\QQ_n$) is when 3- or 4-dimensional Frank copulas with moderate dependence were used in the data generating procedure and no discretization was performed ($k = \infty$). We could not find an explanation for this surprising observation. 

\begin{table}[t!]
\centering
\caption{Percentages of rejection of the null hypothesis of radial symmetry computed from 1000 samples of size $n \in \{50, 100, 200\}$ generated using Procedure~\ref{proc:dgm} with $k \in \{10,50,\infty\}$, $t=1$ and $C$ the 2-, 3- or 4-dimensional exchangeable normal or $t_4$ copula whose bivariate margins have a Kendall's tau of $\tau \in \{0,0.25,0.5,0.75\}$.} 
\label{radSymH0}
\begin{tabular}{rrlrrrrrrrrrrrr}
  \hline
  \multicolumn{3}{c}{} & \multicolumn{6}{c}{normal} & \multicolumn{6}{c}{$t_4$} \\
\cmidrule(lr){4-9} \cmidrule(lr){10-15}
\multicolumn{3}{c}{} & \multicolumn{2}{c}{$d=2$} & \multicolumn{2}{c}{$d=3$} & \multicolumn{2}{c}{$d=4$} & \multicolumn{2}{c}{$d=2$} & \multicolumn{2}{c}{$d=3$} & \multicolumn{2}{c}{$d=4$} \\
\cmidrule(lr){4-5} \cmidrule(lr){6-7} \cmidrule(lr){8-9} \cmidrule(lr){10-11} \cmidrule(lr){12-13}
\cmidrule(lr){14-15} $\tau$ & $n$ & $k$ & $\QQ_n$ & $\QQ_n'$ & $\QQ_n$ & $\QQ_n'$ & $\QQ_n$ & $\QQ_n'$ & $\QQ_n$ & $\QQ_n'$ & $\QQ_n$ & $\QQ_n'$ & $\QQ_n$ & $\QQ_n'$ \\ \hline
0.00 & 50 & $\infty$ & 5.3 & 4.9 & 3.5 & 3.6 & 3.0 & 3.2 & 4.2 & 4.2 & 3.8 & 4.1 & 3.4 & 3.8 \\ 
   &  & 50 & 9.5 & 4.3 & 9.8 & 3.4 & 8.6 & 3.2 & 8.5 & 2.8 & 10.2 & 3.2 & 8.3 & 3.8 \\ 
   &  & 10 & 74.7 & 1.3 & 66.8 & 0.9 & 56.8 & 0.9 & 76.3 & 0.4 & 63.7 & 1.5 & 54.5 & 1.5 \\ 
   & 100 & $\infty$ & 4.3 & 4.2 & 3.7 & 3.4 & 3.8 & 3.8 & 4.5 & 4.1 & 3.7 & 3.6 & 3.3 & 3.3 \\ 
   &  & 50 & 12.4 & 3.6 & 13.5 & 2.5 & 11.2 & 2.5 & 11.9 & 3.0 & 11.4 & 3.3 & 10.5 & 3.3 \\ 
   &  & 10 & 90.4 & 1.5 & 89.3 & 2.2 & 85.7 & 1.3 & 89.5 & 2.1 & 84.3 & 2.2 & 80.0 & 1.7 \\ 
   & 200 & $\infty$ & 4.9 & 5.1 & 3.8 & 3.7 & 3.9 & 3.8 & 4.2 & 4.6 & 4.0 & 4.1 & 4.1 & 4.1 \\ 
   &  & 50 & 21.8 & 3.3 & 20.2 & 2.9 & 17.5 & 3.5 & 17.6 & 3.6 & 16.8 & 3.3 & 16.6 & 4.3 \\ 
   &  & 10 & 94.1 & 3.7 & 94.9 & 2.4 & 94.3 & 1.9 & 93.0 & 3.6 & 93.3 & 3.3 & 92.7 & 2.3 \\ 
  0.25 & 50 & $\infty$ & 4.4 & 4.3 & 3.7 & 3.6 & 3.8 & 3.6 & 4.5 & 4.7 & 4.6 & 4.9 & 3.7 & 4.0 \\ 
   &  & 50 & 8.5 & 2.5 & 9.8 & 2.9 & 10.3 & 3.3 & 9.3 & 2.9 & 8.9 & 3.1 & 8.6 & 2.8 \\ 
   &  & 10 & 76.8 & 0.9 & 73.7 & 1.0 & 69.2 & 1.5 & 74.1 & 0.7 & 70.2 & 1.0 & 62.8 & 1.3 \\ 
   & 100 & $\infty$ & 4.5 & 4.3 & 3.9 & 3.7 & 4.8 & 4.6 & 4.3 & 4.5 & 3.8 & 3.9 & 3.9 & 3.8 \\ 
   &  & 50 & 11.0 & 2.7 & 11.0 & 2.3 & 10.7 & 3.0 & 12.9 & 2.7 & 10.9 & 3.1 & 12.2 & 2.7 \\ 
   &  & 10 & 91.5 & 1.9 & 90.1 & 1.7 & 90.9 & 1.5 & 88.8 & 2.0 & 87.7 & 1.5 & 86.2 & 2.2 \\ 
   & 200 & $\infty$ & 5.0 & 4.8 & 4.0 & 3.8 & 3.9 & 3.8 & 3.7 & 3.8 & 4.6 & 4.6 & 4.0 & 4.3 \\ 
   &  & 50 & 20.8 & 3.4 & 17.7 & 2.6 & 15.4 & 2.7 & 16.1 & 2.6 & 19.3 & 4.0 & 14.1 & 3.0 \\ 
   &  & 10 & 93.1 & 3.9 & 95.1 & 3.2 & 95.1 & 2.5 & 93.0 & 3.7 & 94.3 & 4.0 & 93.0 & 3.3 \\ 
  0.50 & 50 & $\infty$ & 4.3 & 4.4 & 3.7 & 3.6 & 5.1 & 5.0 & 4.9 & 4.8 & 4.1 & 4.3 & 3.8 & 3.8 \\ 
   &  & 50 & 9.5 & 2.4 & 9.0 & 2.8 & 8.1 & 4.1 & 10.5 & 2.8 & 8.5 & 3.0 & 9.3 & 2.7 \\ 
   &  & 10 & 77.4 & 0.6 & 77.3 & 0.5 & 74.8 & 1.0 & 77.0 & 0.8 & 73.7 & 0.9 & 70.9 & 1.0 \\ 
   & 100 & $\infty$ & 4.3 & 4.4 & 4.0 & 4.2 & 5.0 & 4.4 & 4.1 & 4.2 & 4.9 & 4.8 & 3.2 & 3.6 \\ 
   &  & 50 & 11.3 & 2.3 & 9.3 & 3.2 & 8.5 & 3.2 & 11.8 & 2.8 & 11.8 & 3.3 & 9.1 & 2.4 \\ 
   &  & 10 & 92.4 & 0.7 & 90.8 & 1.5 & 91.9 & 1.9 & 90.8 & 1.3 & 92.0 & 1.4 & 91.9 & 1.7 \\ 
   & 200 & $\infty$ & 6.1 & 5.6 & 4.1 & 4.5 & 3.6 & 3.7 & 3.8 & 3.8 & 5.0 & 5.3 & 3.2 & 3.2 \\ 
   &  & 50 & 21.3 & 3.1 & 14.7 & 3.0 & 10.5 & 2.7 & 21.7 & 2.7 & 15.5 & 3.9 & 12.8 & 2.5 \\ 
   &  & 10 & 92.6 & 2.8 & 95.3 & 2.7 & 95.6 & 2.5 & 92.7 & 2.6 & 95.5 & 3.8 & 93.8 & 2.8 \\ 
  0.75 & 50 & $\infty$ & 2.9 & 3.0 & 2.4 & 2.4 & 4.4 & 4.5 & 2.7 & 2.8 & 1.8 & 1.7 & 2.3 & 2.5 \\ 
   &  & 50 & 13.6 & 1.2 & 8.7 & 0.9 & 9.0 & 3.4 & 14.4 & 1.3 & 9.7 & 1.1 & 7.7 & 1.9 \\ 
   &  & 10 & 79.0 & 0.1 & 80.8 & 0.4 & 78.7 & 0.7 & 78.9 & 0.1 & 80.8 & 0.2 & 80.3 & 0.2 \\ 
   & 100 & $\infty$ & 2.9 & 2.8 & 3.4 & 3.4 & 3.8 & 3.9 & 3.4 & 3.5 & 4.1 & 3.5 & 3.2 & 3.4 \\ 
   &  & 50 & 22.7 & 1.2 & 12.2 & 1.6 & 10.5 & 2.3 & 21.6 & 1.6 & 13.6 & 2.2 & 9.9 & 1.8 \\ 
   &  & 10 & 91.4 & 0.9 & 94.3 & 0.9 & 94.4 & 1.1 & 90.6 & 1.2 & 93.4 & 1.1 & 94.8 & 1.2 \\ 
   & 200 & $\infty$ & 3.8 & 3.5 & 3.4 & 3.5 & 3.7 & 3.9 & 4.1 & 3.9 & 3.6 & 3.7 & 2.9 & 3.4 \\ 
   &  & 50 & 54.6 & 1.7 & 25.2 & 1.7 & 17.2 & 2.9 & 54.8 & 1.6 & 22.6 & 2.4 & 18.3 & 1.7 \\ 
   &  & 10 & 91.2 & 3.1 & 93.6 & 2.0 & 93.7 & 2.5 & 91.4 & 2.0 & 92.4 & 3.5 & 93.5 & 2.8 \\ 
   \hline
\end{tabular}
\end{table}

To investigate the power of the test based on $\QQ_n'$, a similar experiment was carried out, the main difference being that non-radially symmetric copulas (the Clayton and the Gumbel--Hougaard) were considered instead. The rejection percentages, reported in Table~\ref{radSymH1}, seem to indicate that the power of the test increases with the dimension $d$ when the amount of ties is small ($k \geq 50$) and, unsurprisingly, that, all other factors being kept fixed, the power decreases as $k$ decreases, that is, as the amount of ties increases.

\begin{table}[t!]
\centering
\caption{Percentages of rejection of the null hypothesis of radial symmetry for the test based on $\QQ_n'$ computed from 1000 samples of size $n \in \{100, 200\}$ generated using Procedure~\ref{proc:dgm} with $k \in \{10,50,\infty\}$, $t=1$ and $C$ the $d$-dimensional Clayton or Gumbel--Hougaard copula with $d \in \{2,3,4\}$ whose bivariate margins have a Kendall's tau of $\tau \in \{0,0.25,0.5,0.75\}$.} 
\label{radSymH1}
\begin{tabular}{rrlrrrrrr}
  \hline
  \multicolumn{3}{c}{} & \multicolumn{3}{c}{Clayton} & \multicolumn{3}{c}{Gumbel--Hougaard} \\ \cmidrule(lr){4-6} \cmidrule(lr){7-9} $\tau$ & $n$ & $k$ & $d=2$ & $d=3$ & $d=4$ & $d=2$ & $d=3$ & $d=4$ \\ \hline
0.00 & 100 & $\infty$ & 4.8 & 4.7 & 3.3 & 4.8 & 4.7 & 3.3 \\ 
   &  & 50 & 3.4 & 3.1 & 2.9 & 3.4 & 3.1 & 2.9 \\ 
   &  & 10 & 1.3 & 1.9 & 2.2 & 1.3 & 1.9 & 2.2 \\ 
   & 200 & $\infty$ & 3.0 & 3.3 & 4.4 & 3.0 & 3.3 & 4.4 \\ 
   &  & 50 & 2.7 & 2.5 & 2.5 & 2.7 & 2.5 & 2.5 \\ 
   &  & 10 & 3.4 & 2.5 & 1.6 & 3.4 & 2.5 & 1.6 \\ 
  0.25 & 100 & $\infty$ & 34.8 & 65.6 & 75.5 & 10.8 & 12.4 & 30.5 \\ 
   &  & 50 & 21.1 & 39.2 & 44.9 & 13.4 & 11.5 & 22.4 \\ 
   &  & 10 & 1.3 & 1.5 & 1.9 & 9.2 & 5.2 & 8.8 \\ 
   & 200 & $\infty$ & 65.4 & 91.9 & 96.7 & 17.6 & 42.5 & 73.4 \\ 
   &  & 50 & 39.5 & 66.7 & 79.1 & 23.3 & 34.5 & 49.5 \\ 
   &  & 10 & 4.3 & 3.0 & 3.5 & 22.3 & 16.4 & 19.9 \\ 
  0.50 & 100 & $\infty$ & 78.2 & 93.9 & 97.6 & 17.0 & 26.4 & 44.7 \\ 
   &  & 50 & 66.6 & 85.8 & 92.2 & 19.6 & 22.6 & 34.7 \\ 
   &  & 10 & 14.5 & 17.0 & 21.4 & 12.4 & 9.0 & 14.3 \\ 
   & 200 & $\infty$ & 98.7 & 100.0 & 100.0 & 37.1 & 65.9 & 85.2 \\ 
   &  & 50 & 96.7 & 99.6 & 100.0 & 41.2 & 58.6 & 69.7 \\ 
   &  & 10 & 38.4 & 40.8 & 39.6 & 39.6 & 29.0 & 27.3 \\ 
  0.75 & 100 & $\infty$ & 90.1 & 98.1 & 98.9 & 13.6 & 17.8 & 29.8 \\ 
   &  & 50 & 84.1 & 94.8 & 96.4 & 14.2 & 13.7 & 21.5 \\ 
   &  & 10 & 28.1 & 32.6 & 35.2 & 7.2 & 4.6 & 5.8 \\ 
   & 200 & $\infty$ & 100.0 & 100.0 & 100.0 & 36.5 & 53.0 & 66.6 \\ 
   &  & 50 & 99.7 & 99.9 & 100.0 & 36.1 & 40.9 & 45.7 \\ 
   &  & 10 & 65.0 & 65.5 & 62.4 & 28.5 & 17.6 & 15.0 \\ 
   \hline
\end{tabular}
\end{table}

\section{Tests of extreme-value dependence}
\label{sec:ev}

Extreme-value copulas are the copulas of random vectors distributed according to the \emph{multivariate extreme-value distribution} \citep[see, e.g,][]{BeiGoeSegTeu04}. The latter distribution may be a natural model when $\bm X_1, \dots, \bm X_n$ are obtained by means of the multivariate extension of the \emph{block maxima method} popularized in the univariate case in the seminal monograph of \cite{Gum58}. As already mentioned in Section~\ref{sec:exch}, a nice overview of characterizations and properties of extreme-value copulas can be found in \cite{GudSeg10}, and related inference procedures are discussed for instance in \cite{BucKoj15}.

When carrying out inference on the unknown copula $C$, extreme-value copulas may also appear outside of the multivariate block maxima framework; see, for instance, Section~\ref{sec:illus}. It is thus of strong interest to be able to assess from $\bm X_1, \dots, \bm X_n$ whether $C$ is an extreme-value copula.

Bivariate tests were proposed by \cite{GhoKhoRiv98}, \cite{BenGenNes09}, \cite{KojYan10c} and \cite{CorGenNes14}, while multivariate tests were considered for instance in \cite{BucDetVol11}, \cite{KojSegYan11}, \cite{Gud12} and \cite{BerBucDet13}. As far as adaptation to ties is concerned, a noticeable exception is the test of \cite{CorGenNes14} which can be adapted to discontinuous margins (see Section~6 in that reference). Apart from the latter test and those of \cite{GhoKhoRiv98} and \cite{BenGenNes09}, all the other tests rely on multiplier-based resampling schemes à la \cite{RemSca09} and it is unclear at this point how they could be adapted to accommodate ties in the coordinate samples of $\bm X_1, \dots, \bm X_n$.

The aim of this section is to propose a version of the bivariate tests of \cite{GhoKhoRiv98} and \cite{BenGenNes09} adapted to ties.

Let $\bm X = (X_1,X_2)$ be a random vector with copula $C$ and continuous marginal c.d.f.s $F_1$ and $F_2$. \cite{GhoKhoRiv98}, and then \cite{BenGenNes09}, studied powerful tests based on the moments of the random variable $W = H(X_1,X_2) = C \{ F_1(X_1), F_2(X_2) \}$, the c.d.f.\ of which, denoted by $K$, is called \emph{Kendall distribution function}. Under the assumption that $C$ is an extreme-value copula, \cite{GhoKhoRiv98} showed that
\begin{align*}
K(t) = \mathrm{P}(W \leq t) = t - (1 - \tau) t \log t, \quad t \in (0,1],
\end{align*}
where $\tau$ is Kendall's tau of $C$, and that, for $k\in \mathbb N$, $\mu_k = E(W^k) = (k \tau + 1)/(k+1)^2$, which for instance implies
that
\begin{align}
-1 + 8 \mu_1 - 9 \mu_2 = 0.\label{eq:test1}
\end{align}
In order to test for extreme-value dependence, \cite{GhoKhoRiv98} suggested to assess whether a sample version of the left-hand side of~\eqref{eq:test1} is significantly different from zero, leading to the statistic
\begin{equation}
\label{eq:Tn}
\TT_n = - 1 + \frac{8}{n(n-1)} \sum_{i \neq j} I_{ij} - \frac{9}{n(n-1)(n-2)} \sum_{i \neq j \neq k} I_{ij} I_{kj},
\end{equation}
where $I_{ij} = \1(X_{i1} \leq X_{j1},X_{i2} \leq X_{j2})$. The latter authors showed that, under extreme-value dependence, $\sqrt{n} \TT_n$ converges in distribution to a centered normal random variable with variance $\sigma^2$. To compute approximate p-values for $\TT_n$, \cite{GhoKhoRiv98} proposed a jackknife estimator of $\sigma^2$. Two alternative estimators were derived more recently by \cite{BenGenNes09}: a finite-sample estimator and an asymptotic one.

In a discussion paper, \cite{GenNesRup11} reported the results of a Monte Carlo experiment showing that the test based on $\TT_n$ does not hold its level when there are ties in the coordinate samples of the available data (similar empirical results will be reported later in this section). In addition, in a preliminary theoretical analysis (see Section 5 in the latter reference), the authors showed that, in the presence of ties, the null asymptotic distribution of $\TT_n$ is not necessarily centered any more.

To adapt the test based on $\TT_n$ to the presence of ties, we conjecture that, under extreme-value dependence, $\TT_n$ follows approximately a normal distribution with mean~$b$ (depending on the marginal empirical c.d.f.s of $\bm X_1,\dots,\bm X_n$) and variance $n^{-1} \sigma^2$, where $\sigma^2$ is the variance appearing in the asymptotic null distribution of $\sqrt{n} \TT_n$ when there are no ties. To carry out the test based on $\TT_n$, we thus propose the following empirical procedure:

\begin{proc}[Approximate p-value for $\TT_n$ in the presence of ties]
\label{proc:ev}
$\strut$
\begin{enumerate}
\item Compute $\TT_n$ using~\eqref{eq:Tn}.
\item Let $\hat \sigma^2$ be an estimate of $\sigma^2$ based either on the jackknife estimator of \cite{GhoKhoRiv98} or on one of the two estimators proposed by \cite{BenGenNes09}.
\item For $j\in\{1,2\}$, compute the average ranks $R_{1j},\dots,R_{nj}$ of $X_{1j},\dots,X_{nj}$ and set $S_{ij} = R_{\sigma_j(i), j}$, $i \in \{1,\dots,n\}$, where $\sigma_j$ is a permutation on $\{1,\dots,n\}$ such that $S_{1j} = R_{\sigma_j(1), j} \leq \dots \leq S_{nj} = R_{\sigma_j(n), j}$.
\item Compute $\hat \tau_b$, Kendall's tau $b$ (the sample version of Kendall's tau corrected for ties), from $\bm X_1,\dots,\bm X_n$.
\item For some moderately large integer $N$, repeat the following steps for every $k \in \{1,\dots,N\}$:
  \begin{enumerate}

  \item Generate a random sample $\bm V_1^{(k)}, \dots, \bm V_n^{(k)}$ from the Gumbel--Hougaard copula with parameter $1 / \{ 1 - \max(\hat \tau_b, 0) \}$.

  \item Set $\pobs{W}_i^{(k)} = \pobs{V}_i^{(k)}$, $i \in \{1,\dots,n\}$, and then, for $j \in \{1,2\}$:
    \begin{itemize}
    \item[-] find a permutation $\rho^{(k)}_j$ on $\{1,\dots,n\}$ such that $\hat W_{\rho^{(k)}_j(1),j}^{(k)} \leq \dots \leq \hat W_{\rho^{(k)}_j(n),j}^{(k)}$,
    \item[-] set $\hat W_{\rho^{(k)}_j(i),j}^{(k)} = \hat V_{\ip{S_{ij}},j}^{(k)}$, $i \in \{1,\dots,n\}$.
    \end{itemize}

\item Let $\TT_n^{(k)}$ stand for the version of $\TT_n$ computed from $\pobs{W}_1^{(k)},\dots,\pobs{W}_n^{(k)}$.
  \end{enumerate}

\item Let $\hat b_N = \frac{1}{N} \sum_{k=1}^N \TT_n^{(k)}$. An approximate p-value for the test is then computed as
$$
2 \Phi \left( - \frac{\sqrt{n} | \TT_n - \hat b_N |}{\hat \sigma} \right),
$$
where $\Phi$ is the c.d.f.\ of the standard normal.
\end{enumerate}
\end{proc}

Let us comment on the previous procedure. The aim of Step~5 is to estimate the unknown bias term $b$. Step~5~(a) generates a random sample from an extreme-value copula whose Kendall's tau is equal to the sample Kendall's tau of $\bm X_1,\dots, \bm X_n$. Because of the presence of ties in the coordinate samples, the latter is naturally taken to be \emph{Kendall's tau $b$}, the sample version of Kendall's tau corrected for ties; see \cite{Ken45} and, for instance, \cite{Agr10}. Step~5~(b) then transforms the generated random sample so that it has the same ``tie structure'' as $\bm X_1,\dots, \bm X_n$. The debatable choice of the Gumbel--Hougaard copula in Step~5~(a) is motivated by the following reasoning:
\begin{enumerate}
\item The numerical experiments carried out in \cite{GenKojNesYan11} indicate that there is hardly any practical difference between the existing bivariate exchangeable one-parameter extreme-value copula families such as the Galambos, Gumbel--Hougaard, H\"usler--Reiss, Tawn or $t$-extreme-value: For a fixed Kendall's tau, these copulas are hard to distinguish numerically.
\item Following the previous point, we conjecture that most bivariate extreme-value copulas can be well approximated by a sufficiently flexible asymmetric extreme-value copula family. Such a family can be constructed using Khoudraji's device. Indeed, if $C_1$ and $C_2$ in~\eqref{eq:khoudraji} are extreme-value copulas, so is $D$ \citep[see, e.g.,][]{GenNes13}.
\item Unfortunately, the fitting of copulas constructed using Khoudraji's device is numerically challenging, which is why we opted for the above imperfect solution that will result in a biased estimate of $b$ under the null in particular when $\bm X_1,\dots,\bm X_n$ are from a distribution with a non-exchangeable copula. As the estimation of copulas constructed from Khoudraji's device becomes more stable numerically, Steps 4 and 5~(a) could be changed accordingly.
\end{enumerate}

In our numerical experiments, we used $N=50$ and considered the three possible aforementioned estimators of $\sigma^2$. The best finite-sample behavior was obtained, overall, using the finite-sample estimator of \citet{BenGenNes09}. For the sake of brevity, we only report the corresponding results. Furthermore, as we continue, to refer to the original test of \cite{BenGenNes09} based on the finite-sample variance estimator, we shall write \emph{the test based on $\TT_n$} while the expression \emph{the test based on $\TT_n'$} will refer to the tie-adapted test based on Procedure~\ref{proc:ev}.

To empirically investigate the levels of the tests, we first generated samples of size $n \in \{50,100,200\}$ from the Gumbel--Hougaard copula with a Kendall's tau of $\tau \in \{0,0.25,0.5,0.75\}$, further discretized as explained in Procedure~\ref{proc:dgm} with $k \in \{10,\infty\}$ and $t = 1$. The rejection percentages are reported in Table~\ref{evH0_GH_all}. As a second experiment, we considered samples generated from non-exchangeable extreme-value copulas constructed using Khoudraji's device. The copula $C_1$ in~\eqref{eq:khoudraji} was taken to be the independence copula, the copula $C_2$ was taken to be the Gumbel--Hougaard with a Kendall's tau of 0.75, the second shape parameter $s_2$ was set to 0.95, while the first shape parameter $s_1$ was taken in $\{0.2,0.4,0.6,0.8\}$. The rejection percentages are reported in Table~\ref{evH0_K_all}. As one can see from Tables~\ref{evH0_GH_all} and~\ref{evH0_K_all}, unlike the test based on $\TT_n$, the one based on $\TT_n'$ does not seem too liberal. Additional results for the test based on $\TT_n'$ for $k \in \{10, 20, 50\}$ and $t \in \{0.5,1,2\}$ are reported in Tables~\ref{evH0_GH_ties} and~\ref{evH0_K_ties} in Appendix~\ref{sec:add_sim_Tn'}. Both tables highlight the fact that it is for the uneven discretization $t=0.5$ that the empirical levels of the test based on $\TT_n'$ are the worst, the test becoming particularly conservative when the discretized samples are generated from asymmetric extreme-value copulas. We conjecture that this is at least partly due to the inadequate estimation, in this case, of the bias term~$b$ in Procedure~\ref{proc:ev}.

\begin{table}[t!]
\centering
\caption{Percentages of rejection of the null hypothesis of extreme-value dependence computed from 1000 samples of size $n \in \{50, 100, 200\}$ generated from the Gumbel--Hougaard copula with a Kendall's tau of $\tau \in \{0,0.25,0.5,0.75\}$ and further discretized as explained in Procedure~\ref{proc:dgm} with $k \in \{10,\infty\}$ and $t = 1$.} 
\label{evH0_GH_all}
\begin{tabular}{rlrrrrrr}
  \hline
  \multicolumn{2}{c}{} & \multicolumn{2}{c}{$n=50$} & \multicolumn{2}{c}{$n=100$} & \multicolumn{2}{c}{$n=200$} \\ \cmidrule(lr){3-4} \cmidrule(lr){5-6} \cmidrule(lr){7-8} $\tau$ & $k$ & $\TT_n$ & $\TT_n'$ & $\TT_n$ & $\TT_n'$ & $\TT_n$ & $\TT_n'$ \\ \hline
0.00 & $\infty$ & 5.9 & 6.6 & 4.3 & 4.4 & 4.4 & 4.5 \\ 
   & 10 & 24.3 & 5.8 & 46.3 & 5.2 & 75.8 & 6.5 \\ 
  0.25 & $\infty$ & 5.6 & 6.3 & 5.5 & 6.2 & 5.3 & 5.7 \\ 
   & 10 & 12.3 & 4.6 & 20.2 & 3.7 & 29.7 & 3.8 \\ 
  0.50 & $\infty$ & 5.8 & 5.2 & 5.1 & 5.1 & 5.2 & 5.2 \\ 
   & 10 & 7.4 & 4.7 & 8.9 & 3.3 & 12.7 & 3.4 \\ 
  0.75 & $\infty$ & 6.9 & 6.5 & 6.4 & 6.4 & 5.6 & 5.4 \\ 
   & 10 & 4.8 & 3.2 & 5.1 & 3.4 & 6.6 & 3.6 \\ 
   \hline
\end{tabular}
\end{table}
\begin{table}[t!]
\centering
\caption{Percentages of rejection of the null hypothesis of extreme-value dependence computed from 1000 samples of size $n \in \{50, 100, 200\}$ generated from~\eqref{eq:khoudraji} with $C_1$ the independence copula, $C_2$ the Gumbel--Hougaard with a Kendall's tau of 0.75, $s_2 = 0.95$ and $s_1 \in \{0.2,0.4,0.6,0.8\}$, further discretized as explained in Procedure~\ref{proc:dgm} with $k \in \{10,\infty\}$ and $t = 1$.} 
\label{evH0_K_all}
\begin{tabular}{rlrrrrrr}
  \hline
  \multicolumn{2}{c}{} & \multicolumn{2}{c}{$n=50$} & \multicolumn{2}{c}{$n=100$} & \multicolumn{2}{c}{$n=200$} \\ \cmidrule(lr){3-4} \cmidrule(lr){5-6} \cmidrule(lr){7-8} $s_1$ & $k$ & $\TT_n$ & $\TT_n'$ & $\TT_n$ & $\TT_n'$ & $\TT_n$ & $\TT_n'$ \\ \hline
0.2 & $\infty$ & 5.1 & 5.2 & 4.4 & 4.3 & 5.6 & 5.1 \\ 
   & 10 & 4.2 & 1.4 & 9.4 & 1.3 & 18.8 & 1.5 \\ 
  0.4 & $\infty$ & 5.0 & 5.6 & 4.7 & 5.1 & 5.0 & 5.3 \\ 
   & 10 & 4.5 & 1.1 & 10.1 & 1.0 & 15.6 & 1.6 \\ 
  0.6 & $\infty$ & 5.8 & 6.2 & 5.7 & 5.7 & 5.8 & 6.4 \\ 
   & 10 & 3.5 & 1.2 & 5.8 & 1.7 & 15.7 & 1.1 \\ 
  0.8 & $\infty$ & 5.4 & 5.7 & 6.4 & 6.2 & 5.0 & 5.5 \\ 
   & 10 & 2.8 & 0.4 & 10.3 & 1.5 & 15.4 & 1.4 \\ 
   \hline
\end{tabular}
\end{table}

As for the tests of exchangeability studied in Section~\ref{sec:exch}, it is of interest to assess whether the proposed adaptation to ties leads to a loss of power when there are no ties. To do so, we generated 1000 samples of size $n \in \{100,200\}$ from a Clayton or Plackett copula with a Kendall's tau of $\tau \in \{0.25,0.5,0.75\}$ and estimated the powers of the tests based on $\TT_n$ and $\TT_n'$. The rejection percentages are reported in Table~\ref{evH1_ex_all}. As one see, the two tests seem equivalent in terms of power. Additional results for the test based on $\TT_n'$ for $k \in \{10, 20, 50\}$ and $t \in \{0.5,1,2\}$ are reported in Table~\ref{evH1_ex_ties}. As expected, the larger~$k$, the more powerful the test, overall. A comparison of Tables~\ref{evH1_ex_all} and~\ref{evH1_ex_ties} reveals that, interestingly enough, the loss of power when switching from $k=\infty$ to $k \in \{10, 20, 50\}$ is not dramatic.

\begin{table}[t!]
\centering
\caption{Percentages of rejection of the null hypothesis of extreme-value dependence computed from 1000 samples of size $n \in \{100, 200\}$ generated from a Clayton or Plackett copula with a Kendall's tau of $\tau \in \{0.25,0.5,0.75\}$.} 
\label{evH1_ex_all}
\begin{tabular}{rrrrrr}
  \hline
  \multicolumn{2}{c}{} & \multicolumn{2}{c}{Clayton} & \multicolumn{2}{c}{Plackett} \\ \cmidrule(lr){3-4} \cmidrule(lr){5-6} $n$ & $\tau$ & $\TT_n$ & $\TT_n'$ & $\TT_n$ & $\TT_n'$ \\ \hline
100 & 0.25 & 37.9 & 38.3 & 16.1 & 15.9 \\ 
   & 0.50 & 69.6 & 69.0 & 20.1 & 20.3 \\ 
   & 0.75 & 61.3 & 60.0 & 11.5 & 12.0 \\ 
  200 & 0.25 & 64.2 & 63.9 & 25.7 & 26.0 \\ 
   & 0.50 & 94.5 & 93.7 & 35.9 & 35.3 \\ 
   & 0.75 & 86.7 & 86.6 & 21.2 & 21.7 \\ 
   \hline
\end{tabular}
\end{table}
\begin{table}[t!]
\centering
\caption{Percentages of rejection of the null hypothesis of extreme-value dependence for the test based on $\TT_n'$ computed from 1000 samples of size $n \in \{100, 200\}$ generated from a Clayton or Plackett copula with a Kendall's tau of 0.5, further discretized as explained in Procedure~\ref{proc:dgm} with $k \in \{10, 20, 50 \}$ and $t \in \{0.5,1,2\}$.} 
\label{evH1_ex_ties}
\begin{tabular}{rrrrrr}
  \hline
  \multicolumn{2}{c}{} & \multicolumn{2}{c}{Clayton} & \multicolumn{2}{c}{Plackett} \\ \cmidrule(lr){3-4} \cmidrule(lr){5-6} $k$ & $t$ & $n=100$ & $n=200$ & $n=100$ & $n=200$ \\ \hline
10 & 1.0 & 60.8 & 90.1 & 15.4 & 26.4 \\ 
   & 2.0 & 51.7 & 79.8 & 9.2 & 16.6 \\ 
   & 0.5 & 31.0 & 66.9 & 5.9 & 13.2 \\ 
  20 & 1.0 & 70.2 & 92.9 & 19.4 & 32.4 \\ 
   & 2.0 & 64.4 & 92.8 & 15.8 & 28.8 \\ 
   & 0.5 & 45.0 & 80.5 & 9.7 & 18.9 \\ 
  50 & 1.0 & 70.6 & 94.4 & 23.7 & 34.9 \\ 
   & 2.0 & 68.5 & 93.1 & 18.7 & 34.6 \\ 
   & 0.5 & 60.3 & 91.3 & 14.7 & 30.6 \\ 
   \hline
\end{tabular}
\end{table}

More results for alternatives involving non-exchangeable copulas are reported in Tables~\ref{evH1_K_all} and~\ref{evH1_K_ties} in Appendix~\ref{sec:add_sim_Tn'}. A comparison between the two tables reveals that, again, the loss of power when switching from $k=\infty$ to $k \in \{10, 20, 50\}$ is not dramatic. From Table~\ref{evH1_K_ties}, we see that it is for the uneven discretizations that the rejection rates are the smallest.

\section{Parametric bootstrap-based goodness-of-fit tests}
\label{sec:gof}

Several parametric bootstrap-based goodness-of-fit tests for copulas were studied in \cite{GenRemBea09}. Their asymptotic null validity can be established using the theoretical results of \cite{GenRem08}. Additional Monte Carlo experiments can be found in \cite{Ber09}. Computationally more efficient versions of these tests that can be shown to be asymptotically equivalent under the null can be obtained by using a resampling scheme based on multipliers \citep[see, e.g.,][]{KojYanHol11,KojYan11,GenHuaDuf13,BerBuc16} instead of a parametric bootstrap.

While it is not clear at the moment how goodness-of-fit tests based on multipliers could be adapted to the presence of ties in the coordinate samples of $\bm X_1,\dots,\bm X_n$, it seems possible to do so for all the parametric bootstrap-based goodness-of-fit tests studied in \cite{GenRemBea09}. To illustrate the necessary additional steps in the resampling procedure, we shall focus on one of the most powerful tests.  Let $\CC = \{C_\theta\}$ be a parametric copula family whose goodness of fit is to be assessed. The tests under consideration are based on the empirical process
\begin{equation}
\label{eq:gof_process}
\Cb_n(\bm u) = \sqrt{n} \{ C_n(\bm u) - C_{\theta_n}(\bm u) \}, \qquad \bm u \in [0,1]^d,
\end{equation}
where $C_n$ is the empirical copula defined in~(\ref{eq:empcop}) and $C_{\theta_n}$ is a parametric estimator of $C$ under the assumption that the unknown copula $C$ belongs to the family $\CC$. The estimator $\theta_n$ of the unknown parameter vector $\theta$ appearing in~\eqref{eq:gof_process} is assumed to be solely based on~\eqref{eq:pobs}. It is the case for instance of the method-of-moments estimator based on the inversion of Kendall's tau \citep[see, e.g.,][and the references therein]{Oak82,Gen87,GenRiv93} or of the maximum pseudo-likelihood estimator of \citet{GenGhoRiv95}.

In the large scale Monte Carlo experiments carried out by \cite{Ber09} and \citet{GenRemBea09}, the statistic
\begin{equation}
\label{eq:Sn}
\SS_n = \int_{[0,1]^d} \Cb_n(\bm u)^2 \dd C_n(\bm u) = \sum_{i=1}^n \{ C_n(\bm{\hat U}_i) - C_{\theta_n}(\bm{\hat U}_i) \}^2
\end{equation}
gave the best results overall. The detailed procedure for computing an approximate p-value for $\SS_n$ when there are no ties is given in Appendix~A of \citet{GenRemBea09}.

Before suggesting a tie-adapted version of the testing procedure, let us make a few remarks:
\begin{itemize}
\item In the presence of ties in the coordinate samples, it is natural to based the method-of-moments estimator involving the inversion of Kendall's tau on its sample version corrected for ties known as Kendall's tau $b$ and already mentioned in Section~\ref{sec:ev}. The latter is not affected by whether maximal or average ranks are used in~\eqref{eq:pobs}.
\item In our simulations involving ties, the maximum pseudo-likelihood estimator of \citet{GenGhoRiv95} was overall found to behave substantially better if average ranks instead of maximal ranks were used in~\eqref{eq:pobs}.
\item Our Monte Carlo experiments also suggest that, with the exception of $\theta_n$, the statistic $\SS_n$ in~\eqref{eq:Sn} should be computed from~\eqref{eq:pobs} based on maximal ranks. 
\end{itemize}

In order to obtain a version of the test based on $\SS_n$ adapted to the presence of ties in the coordinate samples of $\bm X_1,\dots,\bm X_n$, we propose the following procedure:

\begin{proc}[Approximate p-value for $\SS_n$ in the presence of ties]
\label{proc:gof}
$\strut$
\begin{enumerate}
\item For $j\in\{1,\dots,d\}$, compute the average ranks $R_{1j},\dots,R_{nj}$ of $X_{1j},\dots,X_{nj}$ and set $S_{ij} = R_{\sigma_j(i), j}$, $i \in \{1,\dots,n\}$, where $\sigma_j$ is a permutation on $\{1,\dots,n\}$ such that $S_{1j} = R_{\sigma_j(1), j} \leq \dots \leq S_{nj} = R_{\sigma_j(n), j}$.
\item Let $\theta_n$ be the value of one of the two aforementioned rank-based estimators of $\theta$ computed from~\eqref{eq:pobs} using average ranks.
\item With $\theta_n$ from Step~2, compute the test statistic $\SS_n$ defined in~\eqref{eq:Sn} from~\eqref{eq:pobs} using maximal ranks.
\item For some large integer $N$, repeat the following steps for every $k \in \{1,\dots,N\}$:
  \begin{enumerate}
  \item Generate a random sample $\bm V_1^{(k)},\dots,\bm V_n^{(k)}$ from copula $C_{\theta_n}$.

  \item Set $\pobs{W}_i^{(k)} = \bm{V}_i^{(k)}$, $i \in \{1,\dots,n\}$, and then, for $j \in \{1,\dots,d\}$:
    \begin{itemize}
    \item[-] find a permutation $\rho^{(k)}_j$ on $\{1,\dots,n\}$ such that $\hat W_{\rho^{(k)}_j(1),j}^{(k)} \leq \dots \leq \hat W_{\rho^{(k)}_j(n),j}^{(k)}$,
    \item[-] set $\hat W_{\rho^{(k)}_j(i),j}^{(k)} = V_{\ip{S_{ij}},j}^{(k)}$, $i \in \{1,\dots,n\}$.
    \end{itemize}

\item Let
\begin{equation}
\label{eq:psobsk}
\pobs{U}_i^{(k)} = \frac{1}{n+1} (R_{i1}^{(k)},\dots,R_{id}^{(k)}), \qquad i \in \{1,\dots,n\},
\end{equation}
where, for any $j \in \{1,\dots,d\}$, $R_{1j}^{(k)},\dots,R_{nj}^{(k)}$ are the average or maximal ranks computed from $\hat W_{1j}^{(k)},\dots,\hat W_{nj}^{(k)}$.

\item Let $\theta_n^{(k)}$ be the value of the same estimator of $\theta$ as in Step~2 computed from~\eqref{eq:psobsk} using average ranks.

  \item Let
    $$
    C_n^{(k)}(\bm u) = \frac{1}{n} \sum_{i=1}^n \1 (\pobs{U}_i^{(k)} \leq \bm u), \qquad \bm
 u \in [0,1]^d,
    $$
and compute
    $$
    \SS_{n}^{(k)} = \sum_{i=1}^n \{ C_n^{(k)}(\pobs{U}_i^{(k)}) - C_{\theta_n^{(k)}}(\pobs{U}_i^{(k)}) \}^2,
    $$
  where $\pobs{U}_1^{(k)},\dots,\pobs{U}_n^{(k)}$ are computed as in~\eqref{eq:psobsk} using maximal ranks.
  \end{enumerate}
\item An approximate p-value for the test is given by $(N+1)^{-1} \sum_{k=1}^N \{\1(\SS_{n}^{(k)} \geq \SS_n) + 0.5\}$.

\end{enumerate}
\end{proc}

The main differences with the procedure given in Appendix~A of \citet{GenRemBea09} lie in Step~1 and Step~4~(b). The latter step in particular ensures that all the bootstrap samples $\pobs{U}_1^{(k)},\dots,\pobs{U}_n^{(k)}$ in Step~4~(e) have the same marginal univariate empirical c.d.f.s as $\pobs{U}_1,\dots,\pobs{U}_n$ used in Step~3 for computing $\SS_n$. Some thought reveals that, when there are no ties in $\bm X_1,\dots, \bm X_n$, the above tie-adapted procedure leads exactly to the same computations as those in the original procedure of \citet{GenRemBea09}.

As we continue, the expression \emph{the test based on $\SS_n$} will refer to the test based on the above procedure without the key steps for dealing with ties, that is, without Steps~1 and 4~(b) and such that the ranks in~\eqref{eq:psobsk} are directly computed from the components samples of $\bm V_1^{(k)},\dots,\bm V_n^{(k)}$ from Step~4~(a). To designate the tie-adapted test involving all steps in Procedure~\ref{proc:gof}, we will write \emph{the test based on $\SS_n'$}.

Parametric bootstrap-based tests are known to be potentially very computationally expensive as they rely intensively on random number generation and fitting of the hypothesized copula family. For that reason, we restricted our investigation of the finite-sample behaviors of the tests based on $\SS_n$ and $\SS_n'$ to the case of exchangeable bivariate copula families. The parameter $N$ was set to 1000 and the tests were carried out at the 5\% significance level.

We started by investigating the empirical levels of the tests based on $\SS_n$ and $\SS_n'$. As hypothesized copula family $\CC$, we considered the Clayton, Gumbel--Hougaard, Frank and Plackett families. Samples of size $n=150$ were generated using Procedure~\ref{proc:dgm} with $k \in \{10,20\}$, $t \in \{1,2,0.5\}$ and $C \in \CC$ with a Kendall's tau of $\tau \in \{0.25,0.5,0.75\}$. The obtained rejection percentages are reported in Table~\ref{gofH0}. As one can see, the test based on $\SS_n$ (non-adapted to ties) is overall way too liberal when relying on maximum pseudo-likelihood estimation. It seems more robust against ties when based on inversion of Kendall's tau, the empirical levels being nonetheless particularly bad for the uneven discretization corresponding to $t = 0.5$. The (tie-adapted) test based on $\SS_n'$ appears overall well-behaved when involving maximum pseudo-likelihood estimation but seems too liberal in some scenarios when based on inversion of Kendall's tau. The previous observations confirm that Steps~1 and 4~(b) in Procedure~\ref{proc:gof} are crucial in the presence of ties. 

\begin{table}[t!]
\centering
\caption{Percentages of rejection of the null hypothesis that $C \in \CC$ computed from 1000 samples of size $n=150$ generated from $C \in \CC$, where $C$ is equal to the Clayton (Cl), Gumbel--Hougaard (GH), Frank (F) or Plackett (P) copula with a Kendall's tau of $\tau$, further discretized as explained in Procedure~\ref{proc:dgm} with $k \in \{10,20\}$ and $t \in \{1,2,0.5\}$. MPL stands for maximum pseudo-likelihood estimation while, `$\hat \tau_b$' refers to estimation based on inversion of Kendall's tau.} 
\label{gofH0}
\begin{tabular}{lrrrrrrrrrrrrrr}
  \hline
  \multicolumn{3}{c}{} & \multicolumn{4}{c}{$t=1$} & \multicolumn{4}{c}{$t=2$} & \multicolumn{4}{c}{$t=0.5$} \\ \cmidrule(lr){4-7} \cmidrule(lr){8-11} \cmidrule(lr){12-15} \multicolumn{3}{c}{} & \multicolumn{2}{c}{MPL}  & \multicolumn{2}{c}{$\hat \tau_b$} & \multicolumn{2}{c}{MPL}  & \multicolumn{2}{c}{$\hat \tau_b$} & \multicolumn{2}{c}{MPL}  & \multicolumn{2}{c}{$\hat \tau_b$} \\ \cmidrule(lr){4-5} \cmidrule(lr){6-7} \cmidrule(lr){8-9} \cmidrule(lr){10-11} \cmidrule(lr){12-13}\cmidrule(lr){14-15} $\CC$ & $\tau$ & $k$ & $\SS_n$ & $\SS_n'$ & $\SS_n$ & $\SS_n'$ & $\SS_n$ & $\SS_n'$ & $\SS_n$ & $\SS_n'$ & $\SS_n$ & $\SS_n'$ & $\SS_n$ & $\SS_n'$ \\ \hline
Cl & 0.25 & 10 & 20.1 & 5.2 & 5.0 & 5.0 & 2.4 & 5.0 & 2.3 & 5.0 & 100.0 & 8.4 & 15.7 & 5.0 \\ 
   &  & 20 & 4.4 & 4.6 & 5.4 & 5.9 & 2.4 & 5.2 & 4.2 & 6.1 & 100.0 & 6.4 & 13.0 & 5.6 \\ 
   & 0.50 & 10 & 3.4 & 4.5 & 5.3 & 4.9 & 1.7 & 4.6 & 6.1 & 6.5 & 99.8 & 5.1 & 17.4 & 5.0 \\ 
   &  & 20 & 1.7 & 4.5 & 5.0 & 6.5 & 1.9 & 4.3 & 4.0 & 6.3 & 84.9 & 5.5 & 15.4 & 5.6 \\ 
   & 0.75 & 10 & 3.2 & 3.1 & 8.4 & 9.4 & 5.7 & 2.8 & 21.9 & 14.7 & 20.1 & 4.1 & 16.2 & 5.7 \\ 
   &  & 20 & 2.5 & 2.1 & 3.3 & 7.1 & 3.7 & 2.5 & 4.8 & 9.6 & 4.1 & 3.6 & 10.9 & 5.9 \\ 
  GH & 0.25 & 10 & 81.7 & 4.9 & 3.7 & 4.3 & 100.0 & 5.6 & 1.9 & 4.8 & 28.4 & 5.3 & 10.1 & 3.7 \\ 
   &  & 20 & 38.1 & 4.4 & 3.4 & 4.2 & 84.4 & 5.6 & 2.7 & 5.5 & 19.2 & 5.7 & 10.5 & 5.7 \\ 
   & 0.50 & 10 & 83.0 & 4.7 & 3.7 & 4.9 & 100.0 & 5.7 & 2.0 & 6.1 & 29.1 & 6.0 & 14.9 & 5.8 \\ 
   &  & 20 & 34.7 & 4.3 & 3.2 & 5.3 & 91.8 & 6.2 & 2.9 & 5.5 & 20.7 & 4.9 & 11.1 & 5.4 \\ 
   & 0.75 & 10 & 72.2 & 3.7 & 8.1 & 9.3 & 99.7 & 3.4 & 4.7 & 8.6 & 20.8 & 4.8 & 22.4 & 7.2 \\ 
   &  & 20 & 27.7 & 3.9 & 3.4 & 7.2 & 78.1 & 3.5 & 2.8 & 7.1 & 14.3 & 4.1 & 14.5 & 5.6 \\ 
  F & 0.25 & 10 & 5.0 & 4.9 & 4.7 & 4.6 & 4.6 & 4.6 & 1.7 & 4.7 & 96.3 & 5.1 & 12.4 & 4.5 \\ 
   &  & 20 & 5.7 & 5.4 & 4.9 & 5.9 & 5.8 & 5.6 & 3.5 & 5.4 & 43.6 & 4.3 & 8.7 & 4.3 \\ 
   & 0.50 & 10 & 6.0 & 5.3 & 6.5 & 5.9 & 12.5 & 5.9 & 4.1 & 5.8 & 80.6 & 5.2 & 19.7 & 5.4 \\ 
   &  & 20 & 4.9 & 4.6 & 4.1 & 5.1 & 6.8 & 5.3 & 3.3 & 5.2 & 37.7 & 5.6 & 12.7 & 5.8 \\ 
   & 0.75 & 10 & 9.2 & 5.0 & 13.2 & 13.9 & 24.0 & 4.9 & 12.1 & 12.3 & 22.2 & 4.9 & 24.6 & 9.3 \\ 
   &  & 20 & 5.1 & 4.5 & 4.5 & 8.1 & 8.0 & 4.1 & 4.1 & 10.2 & 14.9 & 4.9 & 17.3 & 7.6 \\ 
  P & 0.25 & 10 & 5.3 & 6.0 & 4.5 & 5.2 & 3.6 & 5.1 & 1.9 & 4.9 & 92.5 & 4.6 & 10.2 & 4.4 \\ 
   &  & 20 & 4.8 & 5.1 & 5.0 & 5.8 & 5.7 & 5.2 & 2.7 & 5.7 & 46.5 & 5.3 & 10.4 & 4.9 \\ 
   & 0.50 & 10 & 5.3 & 3.7 & 4.6 & 5.7 & 8.0 & 5.2 & 4.9 & 6.9 & 70.6 & 4.7 & 13.1 & 4.9 \\ 
   &  & 20 & 6.9 & 6.8 & 4.8 & 7.0 & 9.3 & 4.5 & 2.8 & 6.4 & 38.1 & 5.1 & 11.4 & 5.1 \\ 
   & 0.75 & 10 & 9.4 & 3.6 & 7.9 & 10.9 & 9.2 & 4.5 & 6.8 & 10.0 & 22.8 & 8.4 & 17.0 & 6.1 \\ 
   &  & 20 & 6.6 & 4.1 & 4.5 & 8.6 & 11.0 & 3.5 & 3.5 & 8.8 & 9.7 & 5.9 & 10.4 & 6.7 \\ 
   \hline
\end{tabular}
\end{table}

We next decided to focus on the test based on $\SS_n'$ involving maximum pseudo-likelihood estimation and estimated its power in various scenarios under moderate dependence. The results of the numerical experiments are presented in Table~\ref{gof}. Each horizontal block corresponds to a different data generating copula with a Kendall's tau of~0.5. Samples of size $n=150$ were generated using Procedure~\ref{proc:dgm} with $k \in \{10,20,50,\infty\}$ and $t=0.5$. The empirical levels are displayed in bold and confirm the adequate behavior observed previously. When data arise from a copula that is not from the hypothesized family, we see that, unsurprisingly, the power decreases, overall, as $k$ decreases. Finally, one can verify that, as expected, the rejection percentages corresponding to $k=\infty$ are very close to those reported in \citet[Table~2]{GenRemBea09} and \citet[Table~1]{KojYanHol11}.

\begin{table}[t!]
\centering
\caption{Percentages of rejection of the null hypothesis that $C \in \CC$ for the test based on $\SS_n'$ involving maximum pseudo-likelihood estimation computed from 1000 samples of size $n=150$ generated from the Clayton (Cl), Gumbel--Hougaard (GH), Frank (F) or Plackett (P) copula with a Kendall's tau of 0.5, further discretized as explained in Procedure~\ref{proc:dgm} with $k \in \{10,20,50,\infty\}$ and $t=0.5$.} 
\label{gof}
\begin{tabular}{llrrrr}
  \hline
  \multicolumn{2}{c}{} & \multicolumn{4}{c}{Hypothesized copula family $\CC$} \\ \cmidrule(lr){3-6} Data generating copula & $k$ & Cl & GH & F & P \\  \hline
Cl & $\infty$ & \textbf{  4.5} &  99.9 &  89.0 &  84.0 \\ 
   & 50 & \textbf{  4.6} & 100.0 &  90.7 &  89.7 \\ 
   & 20 & \textbf{  4.6} &  99.6 &  76.8 &  78.4 \\ 
   & 10 & \textbf{  4.7} &  98.6 &  57.0 &  53.6 \\ 
   \\GH & $\infty$ & 100.0 & \textbf{  4.3} &  64.3 &  47.1 \\ 
   & 50 & 100.0 & \textbf{  4.6} &  53.5 &  43.7 \\ 
   & 20 &  99.5 & \textbf{  5.7} &  48.9 &  40.0 \\ 
   & 10 &  98.6 & \textbf{  4.1} &  44.2 &  41.3 \\ 
   \\F & $\infty$ &  99.8 &  81.8 & \textbf{  4.0} &  24.4 \\ 
   & 50 &  95.3 &  76.3 & \textbf{  4.5} &  19.0 \\ 
   & 20 &  86.1 &  68.3 & \textbf{  4.9} &  13.1 \\ 
   & 10 &  67.9 &  57.8 & \textbf{  6.1} &   5.8 \\ 
   \\P & $\infty$ &  99.4 &  62.6 &   8.2 & \textbf{  5.4} \\ 
   & 50 &  94.8 &  59.4 &   8.7 & \textbf{  5.2} \\ 
   & 20 &  88.4 &  51.7 &   7.6 & \textbf{  5.2} \\ 
   & 10 &  70.7 &  41.5 &   5.0 & \textbf{  5.0} \\ 
   \hline
\end{tabular}
\end{table}

\section{Illustration}
\label{sec:illus}

To illustrate the use of the studied tie-adapted tests, we considered the LOSS/ALAE insurance data frequently analyzed in the literature \citep[see, e.g.,][]{FreVal98,BenGenNes09,KojYan10,BucKoj15} and available for instance in the \textsf{R} package \texttt{copula}. These consist of 1500 claims of an insurance company on which two variables are observed: LOSS, an indemnity payment, and ALAE, the corresponding allocated loss adjustment expense. We restricted ourselves to the 1466 uncensored claims.  The first component sample, corresponding to the variable LOSS, is particularly affected by ties, with only 541 distinct values, while the second component sample has 1401 distinct values.

We started by applying the tests of exchangeability considered in Section~\ref{sec:exch} with $N=10\,000$ multiplier or bootstrap replicates. The test based on $\RR_{n,C}$ (non-adapted to ties) returned an approximate p-value of 0.000, while the one based on $\RR_{n,C}'$ gave an approximate p-value of 0.049. The tests based on $\RR_{n,A}$ and $\RR_{n,A}'$ returned very similar approximate p-values: 0.119 and 0.112, respectively. As a consequence, we conclude that there is no strong evidence against exchangeability.

Next, we applied the tests of radial symmetry studied in Section~\ref{sec:radSym} with $N=10\,000$ bootstrap replicates. Both the test based on $\QQ_n$ (non-adapted to ties) and the test based on $\QQ_n'$ (adapted to ties) returned an approximate p-value of 0.000, thereby indicating very strong evidence against radial symmetry.

As far as extreme-value dependence is concerned, the test based on $\TT_n$ (non-adapted to ties) returned an approximate p-value of 0.602, while the one based on $\TT_n'$ with $N=1000$ gave a p-value of 0.87. There is therefore no evidence against extreme-value dependence.

\begin{table}[t!]
\centering
\caption{Approximate p-values of the tests studied in Section~\ref{sec:gof} assessing the goodness of fit of the parametric copula families given in column $\CC$ for the LOSS/ALAE data. `$\hat \tau_b$' refers to estimation based on inversion of Kendall's tau, while MPL stands for maximum pseudo-likelihood estimation.} 
\label{loss}
\begin{tabular}{lrrrr}
  \hline
  \multicolumn{1}{c}{} & \multicolumn{2}{c}{$\hat \tau_b$} & \multicolumn{2}{c}{MPL} \\ \cmidrule(lr){2-3} \cmidrule(lr){4-5} $\CC$ & $\SS_n$ & $\SS_n'$ & $\SS_n$ & $\SS_n'$ \\  \hline
GH & 0.230 & 0.221 & 0.192 & 0.167 \\ 
  $\overline{\mathrm{Cl}}$ & 0.000 & 0.000 & 0.000 & 0.000 \\ 
  Cl & 0.000 & 0.000 & 0.000 & 0.000 \\ 
  F & 0.000 & 0.000 & 0.000 & 0.000 \\ 
  N & 0.000 & 0.000 & 0.000 & 0.000 \\ 
  P & 0.000 & 0.000 & 0.000 & 0.000 \\ 
   \hline
\end{tabular}
\end{table}

Finally, we applied the tests studied in Section~\ref{sec:gof} with $N=10\,000$ to assess the goodness of fit of several exchangeable parametric copula families: the Gumbel--Hougaard (GH), the survival Clayton ($\overline{\mathrm{Cl}}$), the Clayton (Cl), the Frank (F), the normal (N) and the Plackett (P). The obtained approximate p-values are reported in Table~\ref{loss}. As one can see, the results of the tests based on $\SS_n$ (non-adapted to ties) and $\SS_n'$ (adapted to ties) are very similar, thereby confirming the robustness against ties of the non-adapted test in certain scenarios as observed in the previous section. Among the six candidate families, only the Gumbel--Hougaard family is not rejected at the 1\% significance level.

\section{Conclusion}

The great bulk of statistical tests for carrying out inference on the copula of a random vector $\bm X$ from i.i.d.\ copies of it were derived under the assumption of continuity of $\bm X$. Many such tests turn out to be too liberal when some coordinate samples of the available data contain ties because of rounding or lack of measurement precision. Ignoring this flaw can obviously lead to a strongly biased inference. As shown in this work, it is however sometimes possible to adapt existing statistical procedures to the presence of ties in the component series of the data: this was carried out for some tests of exchangeability, radial symmetry, extreme-value dependence and goodness of fit. Given the nice properties of multiplier resampling schemes à la \cite{RemSca09} for computing approximate p-values for numerous statistics of interest, a future contribution of strong practical interest would be to find a way to adapt such bootstrap procedures to the presence of ties.

\subsection*{Acknowledgments}

The author is grateful to Johanna G. Ne\v slehov\'{a} for donating the \textsf{R} code implementing the tests studied in \cite{BenGenNes09} to the \texttt{copula} package. Fruitful discussions with Betina Berghaus are also acknowledged.


\bibliographystyle{plainnat}
\bibliography{biblio}

\begin{thebibliography}{56}
\providecommand{\natexlab}[1]{#1}
\providecommand{\url}[1]{\texttt{#1}}
\expandafter\ifx\csname urlstyle\endcsname\relax
  \providecommand{\doi}[1]{doi: #1}\else
  \providecommand{\doi}{doi: \begingroup \urlstyle{rm}\Url}\fi

\bibitem[Agresti(2002)]{Agr02}
A.~Agresti.
\newblock \emph{Categorical Data Analysis}.
\newblock Wiley, second edition, 2002.

\bibitem[Agresti(2010)]{Agr10}
A.~Agresti.
\newblock \emph{Analysis of Ordinal Categorical Data}.
\newblock John Wiley and Sons, New York, second edition, 2010.

\bibitem[Beirlant et~al.(2004)Beirlant, Goegebeur, Segers, and
  Teugels]{BeiGoeSegTeu04}
J.~Beirlant, Y.~Goegebeur, J.~Segers, and J.~Teugels.
\newblock \emph{Statistics of extremes: Theory and Applications}.
\newblock Wiley Series in Probability and Statistics. John Wiley and Sons Ltd.,
  Chichester, 2004.

\bibitem[{Ben Ghorbal} et~al.(2009){Ben Ghorbal}, Genest, and {Ne\v
  slehov\'a}]{BenGenNes09}
M.~{Ben Ghorbal}, C.~Genest, and J.~{Ne\v slehov\'a}.
\newblock On the test of {G}houdi, {K}houdraji, and {R}ivest for extreme-value
  dependence.
\newblock \emph{The Canadian Journal of Statistics}, 37\penalty0 (4):\penalty0
  534--552, 2009.

\bibitem[Berg(2009)]{Ber09}
D.~Berg.
\newblock Copula goodness-of-fit testing: {A}n overview and power comparison.
\newblock \emph{The European Journal of Finance}, 15:\penalty0 675--701, 2009.

\bibitem[Berghaus and B\"ucher(2016)]{BerBuc16}
B.~Berghaus and A.~B\"ucher.
\newblock Goodness-of-fit tests for multivariate copula-based time series
  models.
\newblock \emph{Econometric Theory}, page in press, 2016.

\bibitem[Berghaus et~al.(2013)Berghaus, B\"ucher, and Dette]{BerBucDet13}
B.~Berghaus, A.~B\"ucher, and H.~Dette.
\newblock Minimum distance estimators of the {P}ickands dependence function and
  related tests of multivariate extreme-value dependence.
\newblock \emph{Journal de la Soci\'et\'e Fran\c{c}aise de Statistique},
  154\penalty0 (1):\penalty0 116--137, 2013.

\bibitem[B\"ucher and Kojadinovic(2015)]{BucKoj15}
A.~B\"ucher and I.~Kojadinovic.
\newblock An overview of nonparametric tests of extreme-value dependence and of
  some related statistical procedures.
\newblock In D.~Dey and J.~Yan, editors, \emph{Extreme Value Modeling and Risk
  Analysis: Methods and Applications}, pages 377--398. Chapman and Hall/CRC,
  2015.

\bibitem[B\"ucher et~al.(2011)B\"ucher, Dette, and Volgushev]{BucDetVol11}
A.~B\"ucher, H.~Dette, and S.~Volgushev.
\newblock New estimators of the {P}ickands dependence function and a test for
  extreme-value dependence.
\newblock \emph{The Annals of Statistics}, 39\penalty0 (4):\penalty0
  1963--2006, 2011.

\bibitem[Cap\'era\`a et~al.(1997)Cap\'era\`a, Foug\`eres, and
  Genest]{CapFouGen97}
P.~Cap\'era\`a, A.-L. Foug\`eres, and C.~Genest.
\newblock A nonparametric estimation procedure for bivariate extreme value
  copulas.
\newblock \emph{Biometrika}, 84:\penalty0 567--577, 1997.

\bibitem[Cormier et~al.(2014)Cormier, Genest, and {Ne\v
  slehov\'a}]{CorGenNes14}
E.~Cormier, C.~Genest, and J.G. {Ne\v slehov\'a}.
\newblock Using b-splines for nonparametric inference on bivariate
  extreme-value copulas.
\newblock \emph{Extremes}, 17:\penalty0 633--659, 2014.

\bibitem[Deheuvels(1979)]{Deh79}
P.~Deheuvels.
\newblock La fonction de d\'ependance empirique et ses propri\'et\'es: un test
  non param\'etrique d'ind\'ependance.
\newblock \emph{Acad. Roy. Belg. Bull. Cl. Sci. 5th Ser.}, 65:\penalty0
  274--292, 1979.

\bibitem[Deheuvels(1981)]{Deh81}
P.~Deheuvels.
\newblock A non parametric test for independence.
\newblock \emph{Publications de l'Institut de Statistique de l'Universit\'e de
  Paris}, 26:\penalty0 29--50, 1981.

\bibitem[Fermanian(2013)]{Fer13}
J-D. Fermanian.
\newblock An overview of the goodness-of-fit test problem for copulas.
\newblock In F.~Durante P.~Jaworski and W.~H\"ardle, editors, \emph{Copulae in
  Mathematical and Quantitative Finance}, pages 61--89. Springer, 2013.

\bibitem[Frees and Valdez(1998)]{FreVal98}
E.W. Frees and E.A. Valdez.
\newblock Understanding relationships using copulas.
\newblock \emph{North American Actuarial Journal}, 2:\penalty0 1--25, 1998.

\bibitem[Garralda-Guillem(2000)]{Gar00}
A.I. Garralda-Guillem.
\newblock Structure de d\'ependance des lois de valeurs extr\^emes bivari\'ees.
\newblock \emph{Comptes rendus de l'Acad\'emie des sciences de Paris, S\'erie\
  I Math\'ematique}, 330:\penalty0 593--596, 2000.

\bibitem[Genest(1987)]{Gen87}
C.~Genest.
\newblock Frank's family of bivariate distributions.
\newblock \emph{Biometrika}, 74\penalty0 (3):\penalty0 549--555, 1987.

\bibitem[Genest and {Ne\v slehov\'a}(2007)]{GenNes07}
C.~Genest and J.~{Ne\v slehov\'a}.
\newblock A primer on copulas for count data.
\newblock \emph{The Astin Bulletin}, 37:\penalty0 475--515, 2007.

\bibitem[Genest and {Ne\v slehov\'a}(2013)]{GenNes13}
C.~Genest and J.G. {Ne\v slehov\'a}.
\newblock Assessing and modeling asymmetry in bivariate continuous data.
\newblock In P.~Jaworski, F.~Durante, and W.K. H\"ardle, editors, \emph{Copulae
  in Mathematical and Quantitative Finance}, Lecture Notes in Statistics, pages
  91--114. Springer, 2013.

\bibitem[Genest and Ne\v{s}lehov\'a(2014)]{GenNes14}
C.~Genest and J.~G. Ne\v{s}lehov\'a.
\newblock On tests of radial symmetry for bivariate copulas.
\newblock \emph{Statistical Papers}, 55:\penalty0 1107--1119, 2014.

\bibitem[Genest and R\'emillard(2008)]{GenRem08}
C.~Genest and B.~R\'emillard.
\newblock Validity of the parametric bootstrap for goodness-of-fit testing in
  semiparametric models.
\newblock \emph{Annales de l'Institut Henri Poincar\'e: Probabilit\'es et
  Statistiques}, 44:\penalty0 1096--1127, 2008.

\bibitem[Genest and Rivest(1993)]{GenRiv93}
C.~Genest and L.-P. Rivest.
\newblock Statistical inference procedures for bivariate {A}rchimedean copulas.
\newblock \emph{Journal of the American Statistical Association}, 88\penalty0
  (423):\penalty0 1034--1043, 1993.

\bibitem[Genest and Segers(2009)]{GenSeg09}
C.~Genest and J.~Segers.
\newblock Rank-based inference for bivariate extreme-value copulas.
\newblock \emph{The Annals of Statistics}, 37:\penalty0 2990--3022, 2009.

\bibitem[Genest et~al.(1995)Genest, Ghoudi, and Rivest]{GenGhoRiv95}
C.~Genest, K.~Ghoudi, and L.-P. Rivest.
\newblock A semiparametric estimation procedure of dependence parameters in
  multivariate families of distributions.
\newblock \emph{Biometrika}, 82:\penalty0 543--552, 1995.

\bibitem[Genest et~al.(1998)Genest, Ghoudi, and Rivest]{GenGhoRiv98}
C.~Genest, K.~Ghoudi, and L.-P. Rivest.
\newblock Discussion of ``{U}nderstanding relationships using copulas'', by
  {E}. {F}rees and {E}. {V}aldez.
\newblock \emph{North American Actuarial Journal}, 3:\penalty0 143--149, 1998.

\bibitem[Genest et~al.(2009)Genest, R\'emillard, and Beaudoin]{GenRemBea09}
C.~Genest, B.~R\'emillard, and D.~Beaudoin.
\newblock Goodness-of-fit tests for copulas: {A} review and a power study.
\newblock \emph{Insurance: Mathematics and Economics}, 44:\penalty0 199--213,
  2009.

\bibitem[Genest et~al.(2011{\natexlab{a}})Genest, Kojadinovic, {Ne\v
  slehov\'a}, and Yan]{GenKojNesYan11}
C.~Genest, I.~Kojadinovic, J.~{Ne\v slehov\'a}, and J.~Yan.
\newblock A goodness-of-fit test for bivariate extreme-value copulas.
\newblock \emph{Bernoulli}, 17\penalty0 (1):\penalty0 253--275,
  2011{\natexlab{a}}.

\bibitem[Genest et~al.(2011{\natexlab{b}})Genest, {Ne\v slehov\'a}, and
  Ruppert]{GenNesRup11}
C.~Genest, J.~{Ne\v slehov\'a}, and M.~Ruppert.
\newblock {Comment on the paper by S. Haug, C. Kl\"uppelberg and L. Peng
  entitled ``Statistical models and methods for dependence in insurance
  data''}.
\newblock \emph{Journal of the Korean Statistical Society}, 40:\penalty0
  141--148, 2011{\natexlab{b}}.

\bibitem[Genest et~al.(2012)Genest, {Ne\v slehov\'a}, and Quessy]{GenNesQue12}
C.~Genest, J.~{Ne\v slehov\'a}, and J.-F. Quessy.
\newblock Tests of symmetry for bivariate copulas.
\newblock \emph{Annals of the Institute of Statistical Mathematics},
  64:\penalty0 811--834, 2012.

\bibitem[Genest et~al.(2013)Genest, Huang, and Dufour]{GenHuaDuf13}
C.~Genest, W.~Huang, and J-M. Dufour.
\newblock A regularized goodness-of-fit test for copulas.
\newblock \emph{Journal de la Soci\'et\'e fran\c{c}aise de statistique},
  154:\penalty0 64--77, 2013.

\bibitem[Genest et~al.(2014)Genest, {Ne\v slehov\'a}, and
  R\'emillard]{GenNesRem14}
C.~Genest, J.G. {Ne\v slehov\'a}, and B.~R\'emillard.
\newblock On the empirical multilinear copula process for count data.
\newblock \emph{Bernoulli}, 20:\penalty0 1344--1371, 2014.

\bibitem[Ghoudi et~al.(1998)Ghoudi, Khoudraji, and Rivest]{GhoKhoRiv98}
K.~Ghoudi, A.~Khoudraji, and L.-P. Rivest.
\newblock Propri\'et\'es statistiques des copules de valeurs extr\^emes
  bidimensionnelles.
\newblock \emph{The Canadian Journal of Statistics}, 26\penalty0 (1):\penalty0
  187--197, 1998.

\bibitem[Gudendorf(2012)]{Gud12}
G.~Gudendorf.
\newblock \emph{Nonparametric estimation of multivariate extreme-value
  copulas}.
\newblock PhD thesis, Universit\'e catholique de Louvain, 2012.

\bibitem[Gudendorf and Segers(2010)]{GudSeg10}
G.~Gudendorf and J.~Segers.
\newblock Extreme-value copulas.
\newblock In P.~Jaworski, F.~Durante, W.~H\"ardle, and W.~Rychlik, editors,
  \emph{Copula theory and its applications (Warsaw, 2009)}, Lecture Notes in
  Statistics, pages 127--146. Springer-Verlag, 2010.
\newblock arXiv:0911.1015v2.

\bibitem[Gumbel(1958)]{Gum58}
E.~J. Gumbel.
\newblock \emph{Statistics of extremes}.
\newblock Columbia University Press, New York, 1958.

\bibitem[Hofert et~al.(2017)Hofert, Kojadinovic, M\"achler, and Yan]{copula}
M.~Hofert, I.~Kojadinovic, M.~M\"achler, and J.~Yan.
\newblock \emph{copula: {M}ultivariate dependence with copulas}, 2017.
\newblock URL \url{http://CRAN.R-project.org/package=copula}.
\newblock {R} package version 0.999-17.

\bibitem[Kendall(1945)]{Ken45}
M.~G. Kendall.
\newblock The treatment of ties in rank problems.
\newblock \emph{Biometrika}, 3:\penalty0 239--251, 1945.

\bibitem[Khoudraji(1995)]{Kho95}
A.~Khoudraji.
\newblock \emph{Contributions \`a l'\'etude des copules et \`a la
  mod\'elisation des valeurs extr\^emes bivari\'ees}.
\newblock PhD thesis, Universit\'e Laval, Qu\'ebec, Canada, 1995.

\bibitem[Kojadinovic and Yan(2010{\natexlab{a}})]{KojYan10}
I.~Kojadinovic and J.~Yan.
\newblock Modeling multivariate distributions with continuous margins using the
  \texttt{copula} \textsf{R} package.
\newblock \emph{Journal of Statistical Software}, 34\penalty0 (9):\penalty0
  1--20, 2010{\natexlab{a}}.

\bibitem[Kojadinovic and Yan(2010{\natexlab{b}})]{KojYan10c}
I.~Kojadinovic and J.~Yan.
\newblock Nonparametric rank-based tests of bivariate extreme-value dependence.
\newblock \emph{Journal of Multivariate Analysis}, 101\penalty0 (9):\penalty0
  2234--2249, 2010{\natexlab{b}}.

\bibitem[Kojadinovic and Yan(2011)]{KojYan11}
I.~Kojadinovic and J.~Yan.
\newblock A goodness-of-fit test for multivariate multiparameter copulas based
  on multiplier central limit theorems.
\newblock \emph{Statistics and Computing}, 21\penalty0 (1):\penalty0 17--30,
  2011.

\bibitem[Kojadinovic and Yan(2012)]{KojYan12b}
I.~Kojadinovic and J.~Yan.
\newblock A nonparametric test of exchangeability for extreme-value and
  left-tail decreasing bivariate copulas.
\newblock \emph{Scandinavian Journal of Statistics}, 39\penalty0 (3):\penalty0
  480--496, 2012.

\bibitem[Kojadinovic et~al.(2011{\natexlab{a}})Kojadinovic, Segers, and
  Yan]{KojSegYan11}
I.~Kojadinovic, J.~Segers, and J.~Yan.
\newblock Large-sample tests of extreme-value dependence for multivariate
  copulas.
\newblock \emph{The Canadian Journal of Statistics}, 39\penalty0 (4):\penalty0
  703--720, 2011{\natexlab{a}}.

\bibitem[Kojadinovic et~al.(2011{\natexlab{b}})Kojadinovic, Yan, and
  Holmes]{KojYanHol11}
I.~Kojadinovic, J.~Yan, and M.~Holmes.
\newblock Fast large-sample goodness-of-fit for copulas.
\newblock \emph{Statistica Sinica}, 21\penalty0 (2):\penalty0 841--871,
  2011{\natexlab{b}}.

\bibitem[Liebscher(2008)]{Lie08}
E.~Liebscher.
\newblock Construction of asymmetric multivariate copulas.
\newblock \emph{Journal of Multivariate Analysis}, 99:\penalty0 2234--2250,
  2008.

\bibitem[McNeil et~al.(2015)McNeil, Frey, and Embrechts]{McNFreEmb15}
A.~J. McNeil, R.~Frey, and P.~Embrechts.
\newblock \emph{Quantitative Risk Management: Concepts, Techniques and Tools}.
\newblock Princeton University Press, 2nd edition, 2015.

\bibitem[Nelsen(2006)]{Nel06}
R.B. Nelsen.
\newblock \emph{An introduction to copulas}.
\newblock Springer, New-York, 2006.
\newblock Second edition.

\bibitem[Oakes(1982)]{Oak82}
D.~Oakes.
\newblock A model for association in bivariate survival data.
\newblock \emph{Journal of the Royal Statistical Society Series B},
  44:\penalty0 414--422, 1982.

\bibitem[Pappad{\`a} et~al.(2016)Pappad{\`a}, Durante, and
  Salvadori]{PapDurSal16}
R.~Pappad{\`a}, F.~Durante, and G.~Salvadori.
\newblock Quantification of the environmental structural risk with spoiling
  ties: {I}s randomization worthwhile?
\newblock \emph{Stochastic Environmental Research and Risk Assessment}, 2016.
\newblock URL \url{DOI 10.1007/s00477-016-1357-9}.

\bibitem[Patton(2012)]{Pat12}
A.J. Patton.
\newblock Copula methods for forecasting multivariate time series.
\newblock In \emph{Handbook of Economic Forecasting}, volume~2. Springer
  Verlag, 2012.

\bibitem[Pickands(1981)]{Pic81}
J.~Pickands.
\newblock Multivariate extreme value distributions. {W}ith a discussion.
  {P}roceedings of the 43rd session of the {I}nternatinal {S}tatistical
  {I}nstitute.
\newblock \emph{Bulletin de l'Institut international de statistique},
  49:\penalty0 859--878, 894--902, 1981.

\bibitem[{R Core Team}(2016)]{Rsystem}
{R Core Team}.
\newblock \emph{{R}: {A} Language and Environment for Statistical Computing}.
\newblock R Foundation for Statistical Computing, Vienna, Austria, 2016.
\newblock URL \url{http://www.R-project.org}.

\bibitem[R\'emillard and Scaillet(2009)]{RemSca09}
B.~R\'emillard and O.~Scaillet.
\newblock Testing for equality between two copulas.
\newblock \emph{Journal of Multivariate Analysis}, 100\penalty0 (3):\penalty0
  377--386, 2009.

\bibitem[R{\"u}schendorf(2009)]{Rus09}
L.~R{\"u}schendorf.
\newblock On the distributional transform, {S}klar's {T}heorem, and the
  empirical copula process.
\newblock \emph{Journal of Statistical Planning and Inference}, 139\penalty0
  (11):\penalty0 3921--3927, 2009.

\bibitem[Salvadori et~al.(2007)Salvadori, Michele, Kottegoda, and
  Rosso]{SalDeMKotRos07}
G.~Salvadori, C.~De Michele, N.T. Kottegoda, and R.~Rosso.
\newblock \emph{Extremes in Nature: {A}n Approach Using Copulas}.
\newblock Water Science and Technology Library, Vol. 56. Springer, 2007.

\bibitem[Sklar(1959)]{Skl59}
A.~Sklar.
\newblock Fonctions de r\'epartition \`a $n$ dimensions et leurs marges.
\newblock \emph{Publications de l'Institut de Statistique de l'Universit\'e de
  Paris}, 8:\penalty0 229--231, 1959.

\end{thebibliography}

\appendix

\section{Additional simulations results for the based on $\TT_n'$}
\label{sec:add_sim_Tn'}



\begin{table}[t!]
\centering
\caption{Percentages of rejection of the null hypothesis of extreme-value dependence for the test based on $\TT_n'$ computed from 1000 samples of size $n \in \{50, 100, 200\}$ generated from the Gumbel--Hougaard copula with a Kendall's tau of $\tau \in \{0,0.25,0.5,0.75\}$ and further discretized as explained in Procedure~\ref{proc:dgm} with $k \in \{10,20,50\}$ and $t \in \{0.5,1,2\}$.} 
\label{evH0_GH_ties}
\begin{tabular}{rrrrrr}
  \hline
  $\tau$ & $k$ & $t$ & $n=50$ & $n=100$ & $n=200$ \\ \hline
0.00 & 10 & 1.0 & 6.1 & 5.9 & 4.2 \\ 
   &  & 2.0 & 5.9 & 3.7 & 3.8 \\ 
   &  & 0.5 & 1.4 & 0.9 & 0.6 \\ 
   & 20 & 1.0 & 7.2 & 5.8 & 5.1 \\ 
   &  & 2.0 & 6.7 & 6.7 & 4.6 \\ 
   &  & 0.5 & 3.7 & 2.5 & 1.3 \\ 
   & 50 & 1.0 & 7.5 & 7.8 & 5.4 \\ 
   &  & 2.0 & 6.6 & 6.9 & 6.7 \\ 
   &  & 0.5 & 5.2 & 3.8 & 4.1 \\ 
  0.25 & 10 & 1.0 & 5.2 & 4.4 & 4.2 \\ 
   &  & 2.0 & 4.6 & 2.6 & 3.7 \\ 
   &  & 0.5 & 1.0 & 0.7 & 1.0 \\ 
   & 20 & 1.0 & 6.4 & 6.0 & 5.5 \\ 
   &  & 2.0 & 5.7 & 4.6 & 6.9 \\ 
   &  & 0.5 & 1.7 & 1.9 & 2.3 \\ 
   & 50 & 1.0 & 6.4 & 6.5 & 5.3 \\ 
   &  & 2.0 & 6.8 & 6.1 & 5.1 \\ 
   &  & 0.5 & 4.2 & 3.3 & 4.0 \\ 
  0.50 & 10 & 1.0 & 3.2 & 3.7 & 3.2 \\ 
   &  & 2.0 & 2.8 & 2.8 & 1.8 \\ 
   &  & 0.5 & 1.2 & 0.3 & 1.1 \\ 
   & 20 & 1.0 & 5.4 & 4.8 & 4.0 \\ 
   &  & 2.0 & 4.5 & 4.4 & 3.5 \\ 
   &  & 0.5 & 1.6 & 2.0 & 0.6 \\ 
   & 50 & 1.0 & 6.7 & 5.1 & 5.6 \\ 
   &  & 2.0 & 7.2 & 5.9 & 5.1 \\ 
   &  & 0.5 & 3.8 & 3.3 & 3.0 \\ 
  0.75 & 10 & 1.0 & 2.0 & 2.4 & 2.6 \\ 
   &  & 2.0 & 1.5 & 1.6 & 1.3 \\ 
   &  & 0.5 & 0.9 & 0.7 & 0.4 \\ 
   & 20 & 1.0 & 6.9 & 3.4 & 5.0 \\ 
   &  & 2.0 & 4.8 & 4.1 & 3.5 \\ 
   &  & 0.5 & 2.1 & 0.6 & 0.6 \\ 
   & 50 & 1.0 & 6.6 & 5.9 & 4.7 \\ 
   &  & 2.0 & 6.7 & 4.5 & 5.3 \\ 
   &  & 0.5 & 4.5 & 3.2 & 1.8 \\ 
   \hline
\end{tabular}
\end{table}
\begin{table}[t!]
\centering
\caption{Percentages of rejection of the null hypothesis of extreme-value dependence for the test based on $\TT_n'$ computed from 1000 samples of size $n \in \{50, 100, 200\}$ generated from~\eqref{eq:khoudraji} with $C_1$ the independence copula, $C_2$ the Gumbel--Hougaard with a Kendall's tau of 0.75, $s_2 = 0.95$ and $s_1 \in \{0.2,0.4,0.6,0.8\}$, further discretized as explained in Procedure~\ref{proc:dgm} with $k \in \{10,20,50\}$ and $t \in \{0.5,1,2\}$.} 
\label{evH0_K_ties}
\begin{tabular}{rrrrrr}
  \hline
  $s_1$ & $k$ & $t$ & $n=50$ & $n=100$ & $n=200$ \\ \hline
0.2 & 10 & 1.0 & 1.6 & 1.7 & 2.0 \\ 
   &  & 2.0 & 0.5 & 1.3 & 1.1 \\ 
   &  & 0.5 & 0.1 & 0.1 & 0.1 \\ 
   & 20 & 1.0 & 3.8 & 4.8 & 3.1 \\ 
   &  & 2.0 & 2.6 & 1.8 & 1.9 \\ 
   &  & 0.5 & 0.3 & 0.4 & 0.0 \\ 
   & 50 & 1.0 & 6.1 & 5.5 & 6.0 \\ 
   &  & 2.0 & 4.6 & 3.8 & 4.4 \\ 
   &  & 0.5 & 1.4 & 1.1 & 1.2 \\ 
  0.4 & 10 & 1.0 & 1.1 & 1.2 & 1.6 \\ 
   &  & 2.0 & 0.6 & 0.8 & 0.6 \\ 
   &  & 0.5 & 0.1 & 0.2 & 0.3 \\ 
   & 20 & 1.0 & 3.9 & 2.8 & 3.5 \\ 
   &  & 2.0 & 2.0 & 2.5 & 2.5 \\ 
   &  & 0.5 & 0.8 & 0.4 & 0.3 \\ 
   & 50 & 1.0 & 4.9 & 4.2 & 4.5 \\ 
   &  & 2.0 & 4.5 & 5.3 & 6.3 \\ 
   &  & 0.5 & 2.2 & 1.4 & 1.3 \\ 
  0.6 & 10 & 1.0 & 1.4 & 0.9 & 1.3 \\ 
   &  & 2.0 & 1.1 & 1.0 & 1.1 \\ 
   &  & 0.5 & 0.4 & 0.2 & 0.3 \\ 
   & 20 & 1.0 & 3.6 & 3.2 & 3.3 \\ 
   &  & 2.0 & 3.0 & 2.3 & 1.9 \\ 
   &  & 0.5 & 1.1 & 0.3 & 0.2 \\ 
   & 50 & 1.0 & 4.1 & 5.4 & 5.5 \\ 
   &  & 2.0 & 5.9 & 5.2 & 5.7 \\ 
   &  & 0.5 & 2.1 & 1.3 & 1.2 \\ 
  0.8 & 10 & 1.0 & 0.7 & 1.1 & 2.3 \\ 
   &  & 2.0 & 0.7 & 0.4 & 0.8 \\ 
   &  & 0.5 & 0.2 & 0.1 & 0.2 \\ 
   & 20 & 1.0 & 4.4 & 3.6 & 2.9 \\ 
   &  & 2.0 & 2.6 & 2.5 & 2.6 \\ 
   &  & 0.5 & 0.8 & 0.3 & 0.1 \\ 
   & 50 & 1.0 & 4.7 & 5.7 & 5.1 \\ 
   &  & 2.0 & 4.0 & 3.9 & 5.9 \\ 
   &  & 0.5 & 2.0 & 1.5 & 1.2 \\ 
   \hline
\end{tabular}
\end{table}

\begin{table}[t!]
\centering
\caption{Percentages of rejection of the null hypothesis of extreme-value dependence computed from 1000 samples of size $n \in \{100, 200\}$ generated from~\eqref{eq:khoudraji} with $C_1$ the independence copula, $C_2$ the Clayton or Plackett copula with a Kendall's tau of 0.75, $s_2 = 0.95$ and $s_1 \in \{0.2,0.4,0.6,0.8\}$.} 
\label{evH1_K_all}
\begin{tabular}{rrrrrr}
  \hline
  \multicolumn{2}{c}{} & \multicolumn{2}{c}{Clayton} & \multicolumn{2}{c}{Plackett} \\ \cmidrule(lr){3-4} \cmidrule(lr){5-6} $n$ & $s_1$ & $\TT_n$ & $\TT_n'$ & $\TT_n$ & $\TT_n'$ \\ \hline
100 & 0.2 & 99.9 & 99.9 & 32.9 & 33.6 \\ 
   & 0.4 & 100.0 & 100.0 & 31.2 & 31.1 \\ 
   & 0.6 & 100.0 & 100.0 & 33.3 & 33.2 \\ 
   & 0.8 & 100.0 & 100.0 & 35.6 & 35.6 \\ 
  200 & 0.2 & 100.0 & 100.0 & 61.5 & 61.6 \\ 
   & 0.4 & 100.0 & 100.0 & 58.3 & 58.7 \\ 
   & 0.6 & 100.0 & 100.0 & 60.6 & 61.1 \\ 
   & 0.8 & 100.0 & 100.0 & 62.1 & 61.4 \\ 
   \hline
\end{tabular}
\end{table}
\begin{table}[t!]
\centering
\caption{Percentages of rejection of the null hypothesis of extreme-value dependence based on $\TT_n'$ computed from 1000 samples of size $n \in \{100, 200\}$ generated from~\eqref{eq:khoudraji} with $C_1$ the independence copula, $C_2$ the  Clayton (Cl) or Plackett (P) copula with a Kendall's tau of 0.75, $s_2 = 0.95$ and $s_1 \in \{0.2,0.4,0.6,0.8\}$, further discretized as explained in Procedure~\ref{proc:dgm} with $k \in \{10, 20, 50 \}$ and $t \in \{0.5,1,2\}$.} 
\label{evH1_K_ties}
\begin{tabular}{rrrrrrr}
  \hline
  \multicolumn{3}{c}{} & \multicolumn{2}{c}{$n=100$} & \multicolumn{2}{c}{$n=200$} \\ \cmidrule(lr){4-5} \cmidrule(lr){6-7} $s_1$ & $k$ & $t$ & Cl & P & Cl & P \\ \hline
0.2 & 10 & 1.0 & 99.4 & 15.4 & 100.0 & 35.6 \\ 
   &  & 2.0 & 86.2 & 4.7 & 99.6 & 10.4 \\ 
   &  & 0.5 & 93.4 & 7.7 & 100.0 & 23.4 \\ 
   & 20 & 1.0 & 99.9 & 25.3 & 100.0 & 53.5 \\ 
   &  & 2.0 & 99.6 & 20.6 & 100.0 & 37.9 \\ 
   &  & 0.5 & 97.8 & 12.2 & 100.0 & 29.4 \\ 
   & 50 & 1.0 & 100.0 & 31.6 & 100.0 & 56.9 \\ 
   &  & 2.0 & 100.0 & 26.2 & 100.0 & 54.6 \\ 
   &  & 0.5 & 100.0 & 19.6 & 100.0 & 43.7 \\ 
  0.4 & 10 & 1.0 & 99.2 & 15.4 & 100.0 & 36.7 \\ 
   &  & 2.0 & 86.7 & 4.8 & 99.6 & 11.2 \\ 
   &  & 0.5 & 93.7 & 7.5 & 100.0 & 23.4 \\ 
   & 20 & 1.0 & 99.9 & 27.3 & 100.0 & 52.1 \\ 
   &  & 2.0 & 99.6 & 19.3 & 100.0 & 37.4 \\ 
   &  & 0.5 & 98.0 & 9.7 & 100.0 & 28.8 \\ 
   & 50 & 1.0 & 100.0 & 33.4 & 100.0 & 60.0 \\ 
   &  & 2.0 & 100.0 & 29.2 & 100.0 & 56.1 \\ 
   &  & 0.5 & 99.9 & 19.7 & 100.0 & 44.3 \\ 
  0.6 & 10 & 1.0 & 99.4 & 15.0 & 100.0 & 39.1 \\ 
   &  & 2.0 & 85.7 & 4.2 & 99.7 & 9.9 \\ 
   &  & 0.5 & 93.1 & 8.0 & 100.0 & 22.1 \\ 
   & 20 & 1.0 & 100.0 & 27.8 & 100.0 & 51.3 \\ 
   &  & 2.0 & 99.7 & 19.3 & 100.0 & 39.3 \\ 
   &  & 0.5 & 98.2 & 9.0 & 100.0 & 29.1 \\ 
   & 50 & 1.0 & 100.0 & 29.8 & 100.0 & 58.2 \\ 
   &  & 2.0 & 100.0 & 29.2 & 100.0 & 57.5 \\ 
   &  & 0.5 & 100.0 & 20.9 & 100.0 & 41.5 \\ 
  0.8 & 10 & 1.0 & 98.8 & 14.5 & 100.0 & 37.7 \\ 
   &  & 2.0 & 86.0 & 5.9 & 99.8 & 12.1 \\ 
   &  & 0.5 & 94.8 & 7.8 & 100.0 & 22.7 \\ 
   & 20 & 1.0 & 99.9 & 28.4 & 100.0 & 56.4 \\ 
   &  & 2.0 & 99.6 & 18.2 & 100.0 & 39.7 \\ 
   &  & 0.5 & 97.8 & 11.8 & 100.0 & 31.6 \\ 
   & 50 & 1.0 & 100.0 & 33.9 & 100.0 & 59.4 \\ 
   &  & 2.0 & 100.0 & 26.5 & 100.0 & 54.7 \\ 
   &  & 0.5 & 100.0 & 18.8 & 100.0 & 46.6 \\ 
   \hline
\end{tabular}
\end{table}

\end{document}